 \documentclass[aps,preprint,superscriptaddress,groupedaddress]{revtex4}  
\usepackage{graphicx}  
\usepackage{dcolumn}   
\usepackage{bm}        
\usepackage{amssymb}   
\usepackage{amsmath}
\usepackage{dsfont}
\usepackage{color}

\newcommand{\be}{\begin{equation}}
\newcommand{\ee}{\end{equation}}
\newcommand{\bea}{\begin{eqnarray}}
\newcommand{\eea}{\end{eqnarray}}
\newcommand{\bee}{\begin{eqnarray*}}
\newcommand{\eee}{\end{eqnarray*}}

\newcommand{\nnu}{\nonumber\\}

\newcommand{\mt}{\tilde m}

\newcommand{\at}{\tilde a}

\def\mt{{\ifmmode\td M_t\else $\td M_t$\fi}}
\def\as{{\ifmmode\alpha_s\else$\alpha_s$\fi}}

\let\td=\tilde

\def\co#1{{\ifmmode{\cal O}_{#1}\else${\cal O}_{#1}$\fi}}
\def\cs#1{{\ifmmode{\cal S}_{#1}\else${\cal S}_{#1}$\fi}}
\def\at{{\ifmmode{\tilde A}\else$\tilde A$\fi}}

\def\fr#1.#2.{{#1\over #2}}

\def\mg{{\ifmmode M_{GUT}\else $M_{GUT}$\fi}}

\newcommand{\oot}{\overline {126}}

\usepackage[colorlinks]{hyperref}

\begin{document}

\title{ New minimal supersymmetric GUT emergence and sub-Planckian renormalization group flow}
\author{ Charanjit S. Aulakh\footnote{aulakh@iisermohali.ac.in, aulakh@pu.ac.in}  }

\affiliation{Indian Institute of Science Education and Research
Mohali,\\ Sector 81, S. A. S. Nagar, Manauli PO 140306, India }
\affiliation{International Centre for Theoretical Physics,\\
Strada Costiera 11, 34100,Trieste,  Italy}

 \author{Ila Garg}
\affiliation{Physical Research Laboratory, Ahmedabad- 380009,
India}
\date{\today}

 \author{  Charanjit K. Khosa}
\affiliation{Centre for High Energy Physics, Indian Institute of
Science, Bangalore- 560012, India}
\date{\today}

\begin{abstract} Consistency of trans-unification RG evolution is
used to discuss the domain of definition of the New Minimal Supersymmetric SO(10) GUT
  (NMSGUT). We compute the 1-loop RGE $\beta$ functions, simplifying
  generic  formulae using constraints of gauge invariance
   and superpotential structure. We also calculate the 2 loop contributions
   to the gauge coupling and gaugino mass and indicate how to get full 2 loop
   results for all couplings. Our method overcomes combinatorial barriers that
    frustrate computer algebra based attempts to calculate SO(10) $\beta$
     functions involving large irreps. Use of the RGEs identifies a perturbative
domain $Q < M_E$, where   $M_E <
M_{Planck}$ is  the \emph{scale of emergence} where the NMSGUT, with GUT compatible soft supersymmetry breaking terms emerges from the strong UV dynamics associated
  with the Landau poles in gauge and Yukawa couplings. Due to the strength
  of the RG flows the Landau poles  for gauge and Yukawa couplings
  lie near a cutoff  scale $\Lambda_E $ for the perturbative dynamics of  the NMSGUT which just above $M_E$.   SO(10) RG flows into the IR are shown to 
facilitate small  gaugino masses and generation of negative Non
Universal Higgs masses  squared needed by   realistic NMSGUT
fits of low energy data.   Running the simple canonical theory
emergent at $M_E$ through   $M_X$ down to the electroweak scale
  enables tests of candidate scenarios such as supergravity based NMSGUT with canonical
   kinetic terms and NMSGUT based dynamical Yukawa unification.

\end{abstract}


 \maketitle
 \newpage
\section{ Introduction} \hspace{0.5cm}
Renormalization group equations (RGE) are an important
mathematical tool to study the evolution   of the parameters
(couplings and masses) of a quantum field theory with energy
scale. For example the three gauge couplings of the Standard Model
(SM) evolve with energy and tend to meet roughly around energy
$10^{15}$ GeV : this was the first dynamical  hint  supporting the
``Grand Unification'' vision\cite{patisalam,georgi,gqw}.   However
the SM has a problem in the  sensitivity of the Higgs mass to
quantum effects of superheavy  particles which give rise to large
loop corrections due to their circulation   within loops
correcting the Higgs propagator.  This  implies a mass  correction
: $\Delta m_{H}^2$ $\sim$ $\alpha M_{X}^2$. Supersymmetry (Susy)
is the best known tool to cure this problem. The two loop RGEs of
gauge couplings, superpotential parameters\cite{mahacekvaughn}
and soft terms\cite{jj1,martinvaughn} of a generic softly broken
supersymmetric theory have long been available. The exact relation
between the beta functions for dimensionless and dimensionful
couplings  is also known\cite{JJ2}. In particular these results
give the explicit formulas for the MSSM $\beta$ functions which
are routinely used to study the evolution of MSSM parameters from
UV scales into physically meaningful quantities that describe
physics near the electroweak scale.

The combination of supersymmetry and RG flows 
   leads to nearly exact convergence of the three gauge
couplings  of the MSSM  at $M_X^0=10^{16.3} $ GeV. This striking
and robust result has remained the most convincing  hint of
physics beyond the standard model for nearly 30 years since it was
predicted to be   possible by Marciano and Senjanovic
\cite{marcsenj}  if the top quark mass was found  to be near to
$200$ GeV and $\sin^2\theta_W$ was larger than $0.23$ : as was
found to be the case after  more than a two decades of
 searches and measurements\cite{amaldi}. Apart from the hints from neutrino oscillations
 this amazing convergence has for long  stood as the unique guide post to the nature of
 extreme ultraviolet physics.

  The closeness of the MSSM
  Unification scale to the Planck scale where gravity becomes
  strong has long tantalized theorists. We have   advocated that induced gravity is a natural partner for
   Asymptotically strong GUTs \cite{trmin,tas} since their
    scale of Asymptotic Strength and UV condensation should function both as a UV cutoff for the perturbative GUT and
    set the scale for its contributions to the strength of gravity.  Recent theoretical
  arguments \cite{gia}  renew  the old speculation \cite{adler}  that
  the observed Planck mass may receive dominant or significant contributions $\sim \sqrt{N} M_{GUT}$
   (weakening or inducing gravity   by raising the effective Planck mass from a nominal value $M_P^{0}$  to 
    $M_{P}^{eff}= M_{Pl}^0 + \# \sqrt{N_X} M_{GUT}$)   if there are a large
     number($N_X$)  of heavy particle degrees of freedom
   of mass $M_{GUT}$.   To our mind the most appealing scenario\cite{trmin,tas}
   is the interpretation of the strong coupling scale of the NMSGUT as a physical UV
   momentum cutoff. Simultaneously   gravitational variables(metric,vierbein,gravitino)  are
   demoted    to the role of a  background even as they are supplied with Kinetic terms by the effect of matter
   quantum fluctuations. Their  effective  action and strength
   are  determined by  GUT scale wave function renormalization of the dummy variables
    introduced firstly to implement  general covariance.
   Situational boundary conditions relevant to   large scale
   astrophysical and cosmological  contexts which are, very plausibly,
    the only ones  where gravity is   actually relevant
    will then specify the stress densities that
   source the classical gravitational fields  and waves. Such an acceptance of the
   secondary and induced nature of the gravitational field which needs no
     quantization might finally  lay   quantum gravity to rest  as an  irrelevant
     incubus,  at least to the satisfaction of those concerned with
   testable hypotheses, provided it were anchored in a  interpretation of the Planck cutoff
 as a   physical cutoff arising from the breakdown of GUT perturbativity. Induced gravitational
 kinetic terms have long been postulated\cite{adler} on
   grounds of perturbative wave function renormalization of the graviton due to heavy particles.
 If we take the Planck mass as its experimental value ($10^{18.4}$
 GeV)  and $\Lambda=M_X=10^{16.3}$  GeV this seems to indicate $N\sim 10^4$.
   Thus it is   interesting to note that in
    the NMSGUT there are 640 chiral superfields and 45 Vector superfields.
    This large number of SO(10) coupled fields are precisely
     what make the couplings diverge strongly in the UV.  If we count each chiral
    and vector superfield as 4 degrees of freedom we see that  number of heavy particle degrees of freedom
     is  $N \sim 10^{3.4} $. This is in the right ball park to justify the claim that the
     NMSGUT corrections to the graviton propagator actually reduce gravity
     to the weakly coupled theory we observe. By this line of reasoning the Planck
     scale is  determined by the  unification scale of the NMSGUT  or its
     flavour unifying generalization(the so called
     `YUMGUT'\cite{yumgut} which has even more superfields). In this
     picture  the Landau Pole(s) of the NMSGUT signal a physical cutoff for the perturbative GUT at
     a scale $\Lambda_X\sim 10^{17.0} - 10^{17.5}$ GeV, are the scale of UV condensation driven by
SO(10) gauge forces  and moreover set the observed strength of
gravity.
  Conversely the   observed strength of gravity
actually dictates the precise   value of the UV momentum cutoff
  to be used when computing  GUT quantum effects in any renormalization scheme.
   Thus the relation between different cutoff schemes is presumably
   deducible. However, it must be admitted that there are technical obstacles\cite{davstrom} in
   the way of these largely intuitive arguments which may not only render
   the Newton constant un-calculable but
   necessitate the introduction of the Planck Length as a fundamental parameter and require
     independent  quantization of gravity.

 In the Landau Polar region, the gauge coupling is strong and the
 theory has entered some sort of condensed phase\cite{trmin,tas}.
 Thus the range of scales where the gauge symmetry  of
  the  unified gauge group has unsuppressed play  seems confined to a
  narrow range of scales $ \sim 10^{15.5}< Q< 10^{17.5}$ GeV.
  The UV flows of asymptotically free GUTs (of which, in our opinion,
    no fully realistic example as successful as the realistic Asymptotically
    Strong Susy SO(10) models\cite{nmsgut,bstabhedge} really
  exists) cannot further constrain these  scales  and only seem to offer
  the picture of a weakly coupled gauge theory crushed as an irrelevance  by the
  strength of gravity above $M_{Planck}^{eff}$.  In contrast we
  argue that  asymptotically strong GUTs (ASGUTs)\cite{trmin,tas}  point  to   simple yet phenomenologically
  and calculationally viable   linkage
  between  gravity and Grand Unification of non gravitational
  forces and matter.

  The very asymptotic strength of NMSGUT  RG   flows
  also hints  how the weakly coupled gauge theory  and a weakly coupled gravitational theory
  can emerge  supernatant at large length scales  upon the  condensate of strongly coupled
   physics at the smallest length scales. The IR flows of these theories very
  rapidly drive the coupling from arbitrarily strong coupling  to the   typical 
  values found via RG analyses   near the Supersymmetric Unification scale $g_{10}\sim
  g_5/\sqrt{2} \sim 0.5$ (subscripts 5 and 10 refer to $SU(5)$ and $SO(10) $
  normalizations for the running gauge coupling constant). From
  this point of view   the   trans-unification flows of the
GUT gauge and Yukawa couplings that presumably underwrite the
convergence of MSSM couplings(and third generation Yukawa
couplings\cite{bttau}) at or near $M_X^0\sim 10^{16.3}$ GeV
require  the existence of a regime $ Q< M_E$ where a perturbative
unified theory actually operates as the proper renormalizable
effective theory describing all particle phenomena except gravity.
 The nature of the RG flows in the
trans-unification or sub-Planckian regime has a vital
bearing on many interesting physical questions such as flavour
violation in Susy theories\cite{barbhallstrumia} and the freedom
to choose soft Susy breaking parameters required by realistic fits
beginning from simple and  universal Susy breaking scenarios such
as canonical Supergravity(cSUGRY) type parameters at the upper
limit $M_E$ where the GUT emerges from the strongly coupled UV
regime proper.

The so called New Minimal Supersymmetric SO(10) GUT(\textbf{NMSGUT}) based on
SO(10) gauge group and the ${\bf{210\oplus 126 \oplus \oot \oplus
10 \oplus 120 }}$ Higgs
system\cite{aulmoh,ckn,nmsgut,gutupend,bstabhedge} is the simplest
and most phenomenologically successful ASGUT in existence. It has
repaid thirty years of detailed investigation by exhibiting a
remarkable  flexibility to accommodate  emergent phenomena and
their associated data in one overarching \emph{calculable}
theoretical framework and resolve long outstanding problems of
unification in terms of the quantum effects implied by its
spontaneous symmetry breaking  and associated mass spectra. This
has resulted\cite{nmsgut,bstabhedge} in a realistic unification
model which is compatible with the known data and with distinctive
predictions for the Susy spectra one hopes to observe at the LHC
and/or its successors.  Thus it is now topical to examine the  RG
flows of this theory in the sub-Planckian/trans-unification regime
to see whether they allow consistent definition of a perturbative
GUT over an appreciable energy range.

The NMSGUT requires\cite{nmsgut,bstabhedge} small gaugino masses,
large squark masses and  negative non universal Higgs mass squared
(NUHM) soft parameters to accomplish EW symmetry breaking and  fit
fermion masses. Such parameters require justification, in
particular  for simple cSUGRY scenarios (gravity mediation  with
canonical gauge and scalar kinetic terms).  Soft Susy breaking
parameters in minimal Supergravity (mSUGRY) are typically assumed
to be generated well above the GUT scale   i.e. near the Planck
scale $M_{P} \equiv (8 \pi G_N)^{-1/2}$ = $10^{18.4}$ GeV. To
consider the effect of renormalization from Planckian scales to
GUT scale, when the GUT symmetry is unbroken, one needs the
explicit form of GUT RGEs. As is well known the NMSGUT exhibits a
Landau pole in the generic  UV running of the  gauge
coupling\cite{trmin,tas} quite close to the perturbative scale of
grand unification. In fact   the large coefficients in the $\beta$
functions of  its other couplings imply  the Landau Polar regime
involves all couplings. Thus there can only be a small energy
interval $M_X < E < M_E$ during which the NMSGUT RGEs are
usable. \emph{Due to the strength of the running} it can still
have important effects even over the  short energy range available
in ASGUTs  as compared to the evolution over three decades of
energy in the  flavour violation   study of  SU(5)
SUGRY-GUTs\cite{barbhallstrumia}. If the unification program is
carried out by  running down simple and perturbative data
initially defined at $M_E$ using first the ASGUT RGEs and then the
effective MSSM RGEs(with added   neutrino Seesaw and other exotic
effective operators) then the rapid weakening of ASGUT couplings
towards the IR ensures that a the trans-unification  flow remains
perturbative and the calculation well defined. On the other hand
the UV flow of such theories enters the Landau Polar region just
above $M_E$ implying that we must assume a physical UV cutoff
$\Lambda_E\simeq M_E$  for the whole Grand unification scenario.
Beyond this energy lies the true \emph{cielo  incognito} where all
 couplings are no longer weak : ``Whereof one cannot speak,
one must be silent.''

In spite of their relevance the RGEs for the NMSGUT had so far
never been presented. In principle the application of the generic
formulas of Martin and Vaughn is algorithmic and straightforward.
However Computer Algebra programs\cite{fonseca} that aim to
calculate the RG functions automatically given the Lagrangian
cannot, in practice, handle the combinatorial complexity  in
theories with as many fields as the MSGUT or  NMSGUT.  Using the
vertex structure of the superpotential and SO(10) gauge invariance
as constraints makes the sums over the components of the large
irreps (${\bf{210}}$, ${\bf{126}}$ ,${\bf{\overline{126}}}$ and
${\bf{120}}$) required by the formulas of\cite{martinvaughn}
tractable.  The form of the RGEs for supersymmetric theories is
governed by the supersymmetric non-renormalization theorem
\cite{nonrenomrth} whereby holomorphic(superpotential)  couplings
are free of renormalization except that  arising from
wave-function renormalization.  A similar simplification is
observable in the formulas for the soft couplings and masses. Once
the tricks for computing the one loop anomalous dimensions are
mastered the two loop   anomalous dimensions  and thus $\beta$
functions also follow with some additional combinatorics.
 In this paper we present  the NMSGUT one loop $\beta$ functions.
  However for the case of the gauge coupling and
gaugino mass we also give the two loop results. We have also
calculated the two loop results for the rest of the hard couplings
and soft Susy breaking parameters\cite{phds} and we indicate how
the methods used for the one loop calculation suffice to yield
also the two loop results. The other explicit two loop formulas
and their effect on running will be discussed in a sequel.

 With  strict  assumptions such as
  canonical kinetic terms and canonical supergravity
type soft breaking terms the  gravitino mass parameter ($m_{3/2}$)
and the universal trilinear scalar parameter $A_0$ ($\sim
m_{3/2}$)   are the \emph{only} free parameters since   then there
are not even any  gaugino masses, the common scalar soft
hermitian mass is $m_{3/2}$ and
 the soft bilinear (``B" type) parameters are determined by
 $A_0, m_{3/2}$\cite{weinberg}.
  Then  the  soft
Susy breaking parameters at GUT scale $M_X$   are determined by
running down soft parameters of NMSGUT  from $M_E$ with just these
two soft parameters as input. Of course in general one may also
consider introducing more general soft terms, but our idea here is
to show the power of the SO(10) RG flow to generate suitable soft
terms at $M_X$ even when placed under such strong constraints. The
NMSGUT SSB and effective theory are explicitly calculable in terms
of the fundamental parameters. In practice the extreme non
linearity of  the connection between these parameters and the low
energy data   implies that only a random search procedure (for
parameters defined at $M_E$) combined with RG flows past
intervening thresholds down to $M_Z$ can find acceptable fits of
the SM data. The degree of confidence in the completeness of the
search diminishes exponentially with the increase in  number of
fundamental parameters.  Thus every reduction in the number of
free parameters represents significant progress towards defining a
\emph{falsifiable} model. The  present work may thus be seen as an
attempt not only to improve the UV consistency but also to enforce
a simplification of the fitting problem by using constraints from
the  consistency of trans-unification RG flows.

In fact we shall see that  the SO(10) RG flow will identify an
additional constraint or tuning that must be imposed to keep the
soft holomorphic scalar bilinear (`B') terms  for the MSSM Higgs
pair in the TeV${}^2$ region mandated by  NMSGUT
fits\cite{nmsgut,bstabhedge} as well as a RG flow based scenario
whereby  the values of the B parameters may naturally be left in
this region.   Various seemingly peculiar  aspects of the NMSGUT
parameter  choices may   find an explanation in terms of the RG
flows at high scales. For instance the negative non universal
Higgs mass squared parameters $m^2_{H,\bar{H}}$ which are found in
NMSGUT are also justifiable by the RG flows between $M_E$ and
$M_X$. Minimal SUGRY predicts that all soft scalar masses squared
are positive and equal to $m_{3/2}^2$ at the scale where they are
generated. Soft gaugino masses will be generated at two loops from
the other soft terms but do not arise at one loop if set to zero
to begin with. This justifies the typical hierarchy we observed in
NMSGUT fits whereby sfermions are in the 5-50 TeV range and are
much heavier than the gauginos of the effective MSSM (which lie in
0.2-3 TeV range: depending on the lower limits imposed by hand in
the search). Also the NUHM with negative masses are preferred to
have controlled lepton flavor violation in
Susy-GUTs\cite{LFVSO10}. Similarly choice of the SUGRY emergence
scale below the Planck scale may also allow adjustment of the
gaugino mass and other low energy parameters. The existence of (quasi) fixed points \cite{pendlross,lanzagross,bajcsann}  of 
the RG flow is an important question with a bearing on the physical interpretation of the theory. We have analysed this question for the NMGUT RG equations  but find that neither  fixed nor quasi-fixed points exist.

In Section \textbf{II} we  introduce the formulae of
\cite{martinvaughn} and evaluate them in terms of the parameters
in the  superpotential of the NMSGUT. In Section \textbf{III}  we
present examples of running in the sub-Planckian domain. We discuss the possibility of fixed and quasi-fixed points of the NMSGUT RG flow in Section \textbf{IV}. A 
summary and discussion of  our results is given  in Section \textbf{V}. In the
Appendix we collect the explicit form of  the 1-loop RG $\beta$ functions
of the NMSGUT for soft and hard parameters.

\section{Application of Martin-Vaughn formulae to  the NMSGUT  }
The  generic renormalizable Superpotential  without singlets is
\cite{martinvaughn}  \be W= \frac{1}{6} Y^{ijk}
\Phi_{i}\Phi_{j}\Phi_{k}+\frac{1}{2}\mu^{ij}\Phi_i \Phi_j\ee Here
$\Phi_i$ are  chiral superfields which contain a complex scalars
$\phi_i$ and  Weyl fermions $\psi_i$. The generic collective
indices $i,j,k$   run  over both the different SO(10)  irreps of
the NMSGUT and dimension  of those
  irreps. The generic Lagrangian
corresponding to Soft Susy breaking terms is given by \be L_{Soft
Susy}= -\frac{1}{6} h^{ijk} \phi_{i}\phi_{j}\phi_{k}-\frac{1}{2}
b_{ij}\phi_i
\phi_j-\frac{1}{2}(m^2)^i_j\phi^{*i}\phi_{j}-\frac{1}{2}
M\lambda\lambda +h.c. \ee $ h^{ijk}$ are the soft supersymmetry
breaking  trilinear couplings, $b_{ij}$ the soft breaking bilinear
masses,$(m^2)^i_j$ the Hermitian scalar masses and  $M$ is the
$SO(10)$  gaugino mass parameter. The arrays
 $Y^{ijk},h^{ijk},\mu^{ij},b^{ij}$ are all symmetric and we have
 allowed for  SO(10) invariant universal gaugino masses only
 corresponding to canonical diagonal  gauge kinetic term
 functions and SO(10) invariant 2-loop generation of gaugino masses.

The theory we now call the  New Minimal Supersymmetric SO(10) GUT
 was proposed\cite{aulmoh} by Mohapatra and
one of us (CSA)    in the  early days of supersymmetric GUTs and
was essentially the first complete and consistent supersymmetric
SO(10) GUT. Its natural and minimal structure led another group
\cite{ckn} to independently propose it around the same time.
Following neutrino mass discovery and formulation\cite{MSLRMs} of
high scale left-right and B-L breaking  Minimal Left Right
supersymmetric models in the last years of the last
millenium, it was realized \cite{abmsv} in 2003 that it - and not
another R-parity preserving  supersymmetric SO(10) GUT based on
${\bf{45\oplus 54}}$ \cite{rparso10}, nor any other competing
model such as supersymmetric SU(5) with right handed neutrinos
added - was   the parameter counting minimal realistic Susy SO(10) model. In
the same paper it was shown that the GUT SSB can be reduced to the
solution of a simple cubic equation for one of the vevs.
Thereafter it was the subject of intense study which calculated
its spectra\cite{fuku,bmsv,ag1,ag2} and specified the
roles(coupling magnitudes) required of the different Higgs
representations for complete  fermion
fits\cite{gmblm,bmsvlett,blmdm} taking proper account of the role
of threshold effects at $M_S$ and $M_X$(on gauge unification).
Recently we established\cite{gutupend,bstabhedge} that if proper
account was taken of the threshold effects at $M_X$ on the
relation between effective MSSM and GUT Yukawa couplings then the
latter -which determine fermion masses \emph{and} proton decay-
can emerge so small as to suppress the long standing problematic
fast proton decay due to dimension 5 operators completely.
 The $\bf{210}$, $\bf{126}$ and
$\bf{\overline{126}}$ Higgs   break Susy SO(10) to MSSM. The
$\bf{10}$ and $\bf{120}$ Higgs  are mainly responsible for the
larger charged fermion masses while the  small Yukawa couplings of
$\bf{\overline{126}}$ produce
 adequately large left handed neutrino masses  via the Type I
  seesaw mechanism,  instead of failing to do so due to too large right handed
 neutrino   Majorana masses : as feared from the early days of this
 model\cite{ckn}. Moreover these quantum corrected effective Yukawas restore a
welcome freedom from the onerous constraints (such as
$y_b-y_\tau\simeq y_s-y_\mu,y_b/y_\tau\simeq y_s/y_\mu$
\cite{10plus120}) on fermion Yukawas imposed by the NMSGUT
proposal\cite{blmdm,nmsgut} to use mainly the $\mathbf{10,120}$
irreps for charged fermion masses. Thus the model as it
stands\cite{nmsgut,bstabhedge} is fully realistic
 and invites further scrutiny. The current study is a part of an
 effort to simplify and unify the fitting procedure by matching 
 the MSSM parameters  implied by
  randomly chosen GUT parameters at $M_E$ to the electroweak and
  fermion mass data at $M_Z$  after  RG evolution through,
   and threshold corrections  at,   the intervening scales($M_X,M_S$).

 The  Superpotential of the NMSGUT  is : \bea
W&=&\frac{1}{2}\mu_H H_I^2+ \frac{\mu_{\Phi}}{4!} \Phi_{IJKL}
\Phi_{IJKL} +\frac{\lambda}{4!} \Phi_{IJKL} \Phi_{KLMN}
\Phi_{MNIJ}+ \frac{\mu_{\Sigma}}{5!}\Sigma_{IJKLM}
\overline{\Sigma}_{IJKLM}\nonumber\\&&+\frac{\eta}{4!}\Phi_{IJKL}\Sigma_{IJMNO}
\overline{\Sigma}_{KLMNO}+\frac{1}{4!}H_I \Phi_{JKLM}(\gamma
\Sigma_{IJKLM}+ \overline{\gamma} \overline{\Sigma}_{IJKLM}  )
\nonumber\\&& +\frac{\mu_{\Theta}}{2(3!)}\Theta_{IJK}\Theta_{IJK}
+ \frac{k}{3!}\Theta_{IJK}H_M \Phi_{MIJK}
  +
 \frac{\rho}{4!}\Theta_{IJK}\Theta_{MNK}\Phi_{IJMN} \nonumber\\
  &&+  \frac{1}{2(3!)}
 \Theta_{IJK}\Phi_{KLMN}(\zeta \Sigma_{LMNIJ}
  +  \bar\zeta \bar\Sigma_{LMNIJ})
 +h_{AB} \Psi^T_A C^{(5)}_2 \gamma_I \Psi_B H_I \nonumber \\&& +
\frac{1}{5!}f_{AB} \Psi^T_A C^{(5)}_2 \gamma_{I_1}....\gamma_{I_5}
\Psi_B\overline{\Sigma}_{I_1...I_5} +  \frac{1}{3!}g_{AB} \Psi_A^T
 C_2^{(5)}\gamma_{I_1}...\gamma_{I_3}\Psi_B \Theta_{I_1 I_2
 I_3} \eea
Here middle roman capitals $I,J,K....$ are indices of the vector
of SO(10). All SO(10) tensors are completely antisymmetric and the
5 index ones  also obey duality conditions which  halve their
independent components.   The indices $i,j,k$ of the generic
notation   refer  to both the representation and its internal
(independent) components :  $i \equiv \{R ; r \}; r= 1....d(R) $.
So for example for the \textbf{10}-plet  $i\equiv \{10;I\}$ , but
for the \textbf{45}-plet $i\equiv \{45;[IJ]\}$ with only one
ordering of each  anti-symmetrized pair($I<J$) included. Similarly
for the \textbf{120}-plet the index $i$ will run over all the 120
different combinations of  3 anti-symmetrized  vector indices : $
i\equiv [120; [IJK]: I<J<K] $.

As familiar from the MSSM  the chiral gauge invariants in the
superpotential are the templates for the $SO(10)$ invariant soft
supersymmetry breaking terms.   So corresponding to each term in
the superpotential  we have a soft term in ${\cal{L}}_{SoftSusy}$.
For example we have $\tilde\lambda$ corresponding to $\lambda$,
$b_{\Phi}$ corresponding to $\mu_{\Phi}$  and a Hermitian mass
squared parameter for each Higgs representation. In all we have
$\{ \tilde\lambda, \tilde{k}, \tilde{\rho}, \tilde{\gamma},
\tilde{\bar\gamma}, \tilde{\eta}, \tilde{\zeta},
\tilde{\bar\zeta}$, $\tilde{h}$, $\tilde{f}$, $\tilde{g}$ $\}$,
$\{b_{\Phi}, b_{\Sigma}, b_H, b_{\Theta}\}$ and $\{{ m}^2_{\Phi},
{ m}^2_{\Sigma}, { m}^2_{\bar\Sigma}, { m}^2_{\Theta},$ $ {
m}^2_{H}, {m}^2_{\Psi} \}$ parameters in the NMSGUT soft
Lagrangian,where ${m}^2_{\Psi}$ is a 3 by 3 hermitian matrix.

 Our successful fits \cite{nmsgut,bstabhedge} show  that fermion
data  and   EW symmetry breaking  requires negative Higgs soft
masses $m^2_{H,\overline H} $  and soft parameter $b_H$ both negative with magnitudes $\sim \, 10^{10}$ GeV${}^2$ in  ($b_H$ runs positive at low scales)
along with gaugino masses in the TeV(gluino) and
sub-TeV(Bino,Wino) range. In the following sections we will see
that such initial values of the soft parameters can be generated
by running of the SO(10) theory specified above even over the
short range   from $M_E$ to $M_X =M_{GUT}$ and even when beginning
from very restricted scenarios for the initial parameter values :
such as those implied by cSUGRY.

We define the $\beta$ functions at $n$-loop order for any
parameter $x$ after  extracting $n$ powers of $1/(16 \pi^2)$  for
convenience in presentation : \be
\frac{dx}{dt}=\sum_{n=1}\frac{\beta_x^{(n)}}{(16 \pi^2)^n} \ee

 {- The one-loop
$\beta$-functions for the SO(10)
 gauge coupling and gaugino mass parameter M have the generic  form:
 \be
\beta_g^{(1)}=g^3[S(R)-3 C(G)]\quad\quad;\quad\quad
\beta_M^{(1)}=2\beta_g^{(1)} M/g\ee here $S(R)$ and $C(G)$ are
Dynkin index (including contribution of all superfields) and
Casimir invariant respectively.   Table \ref{Dynkin} gives the
Dynkin index and Casimir invariant for different representations
of NMSGUT. We get a total index S(R)=1+(3$\times$
2)+28+35+35+56=161.
\begin{table}[htb]
 $$
 \begin{array}{ccc}
 \hline\hline{\mbox{d}}&{\rm S(R)} &{\rm C(R)=d(G) S(R)/d(R)} \\
 \hline\hline
 45 & 8 & 8  \\
 10 & 1 & 9/2 \\
 16 & 2 & 45/8 \\
 120 & 28 & 21/2\\
 126 & 35 & 25/2\\
 \overline{126} & 35 & 25/2\\
 210 & 56 & 12 \\
 \hline\hline
  \end{array}
 $$
\caption{Dynkin index and Casimir invariant for different
representations of NMSGUT}
 \label{Dynkin} \end{table}
So one-loop $\beta$ functions for the SO(10) gauge coupling and
gaugino mass parameter are \bea
 \beta_{g_{10}}^{(1)}=137g_{10}^3 \label{betag1loop}\\
 \beta_{M}^{(1)}= 274 M g_{10}^2\eea

The general form of 1-loop beta function for Yukawa couplings is
  \cite{martinvaughn}
  : \be{
[\beta_Y^{(1)}]^{ijk}=Y^{ijp} \gamma_p^{(1)k}+(k\leftrightarrow i
)+(k \leftrightarrow j ) }\ee where $\gamma^{(1)}$ is the one loop
anomalous dimension matrix. Thus we  need to calculate the
anomalous dimensions for each superfield.
     SO(10) gauge invariance implies that $\gamma^i_j$ must be field-wise and
(irrep)  componentwise diagonal. This simplifies their computation
enormously. The generic one-loop anomalous dimension parameters
are given by \be { \gamma_i^{(1)j}=\bar{\gamma}_i^{(1)j}-2
g^2 \delta_i^j C(i) }\qquad ; \qquad
\bar{\gamma}_i^{(1)j}\equiv\frac{1}{2} Y_{ipq}Y^{jpq} \label{gammai}\ee
with $Y_{ijk}\equiv Y^{{ijk}*}$.

To see what is involved in calculating $\bar{\gamma}_i^{(1)j}$
consider the example of the \textbf{210}-plet. The independent
components of this irrep correspond to non identical combinations
of four ordered and unequal vector indices : $I<J<K<L$. Let us
select one say 1234. SO(10) invariance requires that
$\bar{\gamma}_i^j$  is diagonal so that   $i\equiv{1234}$ requires
$ j\equiv 1234$ : the propagator correction will obviously not
allow mixing with a different representation than \textbf{210}. So
we are required to sum over all possible symmetric combinations of
independent 210 components  $pq\equiv (\{I<J<K<L\},\{I'<J'<K'<L'\}
)$.

To calculate $\gamma_\Phi^{(1)}$ we must therefore  :
\begin{itemize}
 \item Identify the  combinations of the chosen component(1234)  of $\Phi$
  with other superfields of the model in trilinear gauge invariants.
 \item For any given coupling vertex, calculate the number of ways
  the (conserved) chosen (1234) line  gets wave function corrections from the fields it
   couples to in the considered  vertex.  Since it must emerge with the same SO(10) quantum numbers as it
    entered with and the counting will apply equally
   to every such field component, a little practice suffices to get
    all 1-loop anomalous dimensions.
 \end{itemize}

  Consider  first the   coupling   $\rho \, \Phi_{IJKL} \Theta_{IJM} \Theta_{KLM}
  $
\[ \frac{\rho}{4!}  \Phi_{IJKL} \Theta_{IJM} \Theta_{KLM} =
\sum\limits_{M} \frac{\rho}{4!} 4.2.\Phi_{1234} (\Theta_{12M}
\Theta_{34M}-\Theta_{13M} \Theta_{24M}+\Theta_{14M} \Theta_{23M} )
\] Here M runs over remaining 6 values ($M=5,6..10$ since the $\bf{120}$ plet
is totally antisymmetric).    In this example   we can have 18
possible combinations that couple to $\Phi_{1234}$. Therefore the
contribution to $(\gamma^{(1)})_{1234}^{1234}$ is
  \be  \frac{1}{2}\bigg|Y^{\{\Phi_{1234}.\Theta.\Theta\}}\bigg|^2=\frac{18|\rho|^2}{9}=2 |\rho|^2 \ee

Similarly \be \frac{\gamma}{4!} \Phi_{IJKL}H_M \Sigma_{IJKLM} =
\gamma \Phi_{1234} (H_5 \Sigma_{12345}+H_6 \Sigma_{12346}+....)
\ee The six allowed index values for $H$ (i.e. 5-10) give-in an
obvious shorthand with SO(10)
indices suppressed- \be \sum_{H,\Sigma} Y_{\{\Phi_{1234}.H.\Sigma
\}} Y^{\{\Phi_{1234}.H.\Sigma\}}=6|\gamma|^2 \ee The invariant $k
H_I \Theta_{JKL} \Phi_{IJKL}$ will contribute to
$\gamma_\Phi^{(1)}$ \be \frac{k}{3!} H_I \Theta_{JKL} \Phi_{IJKL}
= k \Phi_{1234} (H_1 \Theta_{234}-H_2 \Theta_{134}+H_3
\Theta_{124}-H_4 \Theta_{312} )+.... \ee \be \sum_{H,\Theta}
Y_{\{\Phi_{1234}.H.\Theta \}} Y^{\{\Phi_{1234}.H.\Theta\}}=4|k|^2
\ee Thus the anomalous dimension matrix reduces to a common
anomalous dimension for each independent component of each field
and   only for the triplicated matter \textbf{16}-plets need one
consider mixing.

In this way one finds that the one  loop anomalous dimension for
the $\mathbf{210}$-plet $\Phi$   is
 \be \gamma_\Phi^{(1)}=4 |k
|^2+180 |\lambda|^2+2 |\rho |^2+240 |\eta |^2+6
(|\gamma|^2+|\bar{\gamma}|^2)+60 (|\zeta |^2+|\bar{\zeta}|^2)-24
g_{10}^2 \label{gammaPhi1}\ee
 Using  the anomalous dimensions  one can
compute the beta functions  for all the superpotential parameters.
For example the  one loop $\beta$ function for $\lambda$ is : \be
\quad \quad \beta_\lambda^{(1)}=  3 \gamma_\Phi^{(1)}  \lambda \ee
 The formulas for the soft terms are closely analogous to those
 for the Superpotential couplings on which they are modelled.  Indeed the exact prescription for obtaining    the  soft from  hard beta functions is known\cite{JJ2}  in terms of a differential operator in the couplings operating on the anomalous dimensions.  This yields the generic  formulae given in \cite{JJ2,martinvaughn}
 
 \bea [\beta^{(1)}_h]^{ijk}&=&
\frac{1}{2}h^{ijl}Y_{lmn}Y^{mnk}+Y^{ijl}Y_{lmn}h^{mnk}-2(h^{ijk}-2M
Y^{ijk})g^2 C(k) \nonumber\\&&
 +(k \longleftrightarrow i)+(k \longleftrightarrow j)\nnu
&=& h^{ijl} \bar\gamma^{(1)k}_l   + 2  Y^{ijl}
\tilde\gamma^{(1)k}_l -2(h^{ijk}-2M Y^{ijk})g^2 C(k) \nonumber\\&&
 +(k \longleftrightarrow i)+(k \longleftrightarrow j)\nnu
 \label{1looph}\eea
where  \be  \tilde\gamma^{(1)k}_l\equiv {\frac{1}{2}} Y_{lmn}
h_{mnk}\ee The index patterns of the soft and hard couplings being
identical one can calculate the one-loop $\beta$ function for the soft
parameter $\tilde{\lambda}$ using the same counting rules used
above to sum over independent loops. For example the $\beta$
function for the soft trilinear  analog of the 210 cubic
superpotential coupling $\lambda$(called $\tilde \lambda$)  is
given by  : \be \beta_{\tilde{\lambda}}^{(1)}=3 \tilde{\lambda}
\bar{\gamma}_{\Phi}^{(1)}+6 \lambda \tilde{\gamma}_{\Phi}^{(1)}
-72g_{10}^2(\tilde{\lambda}-2 M \lambda) \ee where $\bar
\gamma_\Phi^{(1)}$ = $ \frac{1}{2} Y_{\Phi mn}Y^{mn\Phi}$ and
$\tilde{ \gamma}_\Phi^{(1)}$ = $\frac{1}{2} Y_{\Phi mn}h^{mn\Phi}$
are anomalous dimensions. The first was given above in
eqn(\ref{gammaPhi1}) while its soft (tilde) counterpart is \be
\tilde{ \gamma}_\Phi^{(1)}=  4 \tilde{\kappa} \kappa^*+180
   \tilde{\lambda} \lambda ^{*}+2
\tilde{\rho } \rho ^{*} +240
   \tilde{\eta } \eta ^{*} +6
   (\tilde{\gamma}  \gamma
   ^{*}+\tilde{\bar{\gamma}}
   \bar\gamma^{*})+60
   (\tilde{\zeta}\zeta
   ^{*}+\tilde{\bar{\zeta}}
   \bar \zeta ^{*})    \label{tilgamphi}   \ee
These are calculated in  the  way described earlier with
substitution of a soft coupling (h) for a hard coupling (Y)(on
which h is modelled)  and thus the numerical coefficients follow
$\bar\gamma_\Phi $ closely.

The generic form of the $\beta$ functions for the soft bilinear
``B" terms is also known in terms of an exact relation given by the action of a differential operator in the couplings acting on the anomalous dimensions  \cite{JJ2}  and can be found in \cite{JJ2,martinvaughn}  \bea [\beta^{(1)}_b ]^{ij} & =&
 {\frac{1}{ 2}}b^{il} Y_{lmn} Y^{mnj} +{\frac{1}{ 2}}Y^{ijl} Y_{lmn} b^{mn}
+ \mu^{il} Y_{lmn} h^{mnj} \nonumber\\&&- 2  (b^{ij} - 2 M
\mu^{ij}  )g^2 C(i)+ (i \leftrightarrow j)  \eea Which can again
be written in terms of $\bar\gamma$ and $\tilde\gamma$ . Then
arguments similar to those given above yield : \be
\beta_{b_{\Phi}}^{(1)}=2 b_{\Phi}
\bar{\gamma}_{\Phi}^{(1)}+4\mu_{\Phi}
\tilde{\gamma}_{\Phi}^{(1)}-48 g_{10}^2(b_{\Phi}-2 {M}
\mu_{\Phi})\ee Similarly the Hermitian soft masses have generic
$\beta$ functions \bea [\beta^{(1)}_{m^2} ]_i^j &= &
 {\frac{1}{ 2}} Y_{ipq} Y^{pqn} {(m^2)}_n^j
+{\frac{1}{ 2}} Y^{jpq} Y_{pqn} {(m^2)}_i^n + 2 Y_{ipq} Y^{jpr}
{(m^2)}_r^q \nonumber\\&&+ h_{ipq} h^{jpq}- 8\delta_i^j M
M^\dagger g^2 C(i) +
 2g^2{\bf t}^{Aj}_i {\rm Tr}[{\bf t}^A m^2 ]  \eea
Again the previous results and a similar one for the doubly soft
contribution (i.e. from $h^{jpq}h_{ipq}$)  yields for example for
the \textbf{210}  soft Hermitian mass : \bea
\beta_{{m}^2_{\Phi}}^{(1)}&=&2
\bar{\gamma}_{\Phi}^{(1)}{m}^2_{\Phi}+720 {m}^2_{\Phi}|\lambda|^2+
{m}^2_{H}(12|\gamma|^2+12|\bar{\gamma}|^2+8|k|^2)\nonumber\\&&
+{m}^2_{\Theta}(8|\rho|^2+120(|\zeta|^2+|\bar{\zeta}|^2)+8|k|^2)
+{m}^2_{\Sigma}(480|\eta|^2+12|\gamma|^2+120|\zeta|^2)\nonumber\\&&+{m}^2_{\bar{\Sigma}}(480|\eta|^2
+12|\bar{\gamma}|^2+120|\bar{\zeta}|^2)
+2\hat{\gamma}_{\Phi}^{(1)}-96|{M}|^2 g_{10}^2\eea Where
\be\hat\gamma_i^{(1)j}=\frac{1}{2} h_{ipq}h^{jpq} \ee Thus for
example  \be
\hat{\gamma}_{\Phi}^{(1)}=240|\tilde{\eta}|^2+4|\tilde{\kappa}|^2+180|\tilde{\lambda}|^2+2|\tilde{\rho}|^2+
6(|\tilde{\gamma}|^2+|\tilde{\bar{\gamma}}|^2)+60(|\tilde{\zeta}|^2+|\tilde{\bar{\zeta}}|^2)\ee
As a final example of one loop functions  consider matter field
($\Psi_A$) wave function renormalization due to  the matter Higgs
superpotential couplings \be W= h_{AB} \Psi_{A\alpha}
(C\Gamma_I)_{\alpha\beta}\Psi_{B\beta} H_I \ee where  the SO(10)
conjugation matrix $C$  and Gamma matrices $ \Gamma_I$ may be
found in \cite{ag1}, $\alpha,\beta$ are Spin(10) spinor indices
and $A,B..$ are the generation indices. To calculate the
contribution to wavefunction renormalization  we need to contract
 this vertex and its conjugate so as to leave $\Psi_{A
\alpha}$,$\Psi_{A' \alpha'}^*$  as external fields. The remaining
numerical factors are   : \be \sum_{B ,\beta } h_{AB} h^*_{A'B}
(C\Gamma_I)_{\alpha\beta}
(C\Gamma_I)^*_{\alpha'\beta}=(C^*\Gamma_I^* \Gamma^T_I
C^T)_{\alpha' \alpha} (h^*h^T)_{A'A} \ee Then\cite{ag1} either
$C=C_1^{(5)}\equiv \tau_1\times \epsilon \times \tau_1\times
\epsilon \times \tau_1  $ or $C=C_2^{(5)}\equiv \epsilon \times
\tau_1\times \epsilon \times \tau_1\times \epsilon $    and
$\Gamma_i=\Gamma_i^{\dagger}$ easily give $10 \delta_{\alpha'
\alpha} (h^*h^T)_{A'A} $. Similarly the 120 plet contributes  $120
(g^* g^T)$ while the $\mathbf{126-\oot}$ pair give $252(f^* f^T)$
(since there is a   double counting of the 126-plet  components
due to  duality within the 252 independent antisymmetric orderings
of 5 vector indices). Finally since $h,f,g$ are either symmetric
or anti-symmetric $h^*h^T \equiv h^\dagger h, g^*g^T\equiv
g^\dagger g $ etc. The complete 1-loop anomalous dimensions  and  $\beta$ functions are
given in Appendix.

\subsection{Two loop anomalous dimensions}
In this paper we study  RG flows at one loop level with two
important exceptions. Firstly the gauge coupling is strongly
driven to a Landau pole and it is natural to first ask what is the two
loop correction to the huge positive coefficient in the one loop
term. The generic two loop formula is \bea
\beta_g^{(2)}=g^5\big\{-6[C(G)]^2 + 2 C(G) S(R) + 4 S(R)
C(R)\big\}- g^3 Y^{ijk}Y_{ijk} C(k)/d(G)
\label{gauge2loopbeta}\eea

where the factor in the last term simplifies as
$C(k)/d(G)=S(k)/d(k) $. Since for any given field type $k$
$\sum_{ij}Y^{ijk}Y_{ijk}$ is diagonal in field type it follows
that the sum over $k$ will just cancel the dimension of the
representation ($d(k)$) leaving the index S(k) as an overall
factor weighting the contribution of that field type in the last term in eqn. (\ref{gauge2loopbeta}). This yields
\bea  \beta_{g_{10}}^{(2)}&=&  9709 g^5_{10}- {2 g^3_{10}} (
\bar\gamma_{H}^{(1)}+28\bar\gamma_{\Theta}^{(1)}+
35\bar\gamma_{\Sigma}^{(1)}+35\bar\gamma_{\bar\Sigma}^{(1)}
+56\bar\gamma_{\Phi}^{(1)}\nonumber\\&&+2
Tr[\bar\gamma_\psi^{(1)}])  \eea The general formula for the two
loop gaugino mass $\beta$ function is very similar to the gauge
beta function \bea \beta_M^{(2)}&=& 4 g^4\big\{-6[C(g)]^2 + 2 C(G)
S(R) + 4 S(R) C(R)\big\}M+ 2 g^2 (h^{ijk}-M Y^{ijk}) Y_{ijk})
C(k)/d(G) \label{gaugeinomass2loopbeta}\nnu \eea and this readily
evaluates to \bea  \beta_{ M}^{(2)}&=&  38836 g^4_{10}  M+ {4
g^2_{10}} ( (\tilde\gamma_{H}^{(1)}- M\bar\gamma_{H}^{(1)})+2
Tr[\tilde\gamma_\Psi -M\bar\gamma_\Psi] +
28(\tilde\gamma_{\Theta}^{(1)}- M \bar\gamma_{\Theta}^{(1)})+ \nnu
&&35(\tilde\gamma_{\Sigma}^{(1)}-
M\bar\gamma_{\Sigma}^{(1)})\nonumber+35(\tilde\gamma_{\bar
\Sigma}^{(1)}-
M\bar\gamma_{\bar\Sigma}^{(1)})+56(\tilde\gamma_{\Phi}^{(1)}-
 M \bar\gamma_{\Phi}^{(1)}) ) \eea This concludes the
$\beta$ equations we need in this paper. However we have also
computed the complete two loop results \cite{phds}. Here we
indicate how they are computed. The two loop anomalous dimensions
$\gamma^{(2)}$ are the building blocks of two loop $\beta$
functions  and  have generic form : \bea
\gamma_i^{(2)j}&=&-\frac{1}{2}Y_{imn}Y^{npq}Y_{pqr}Y^{mrj}+g^2_{10}Y_{ipq}Y^{jpq}[2C(p)-C(i)]\nonumber\\&&+
2 \delta_i^j g^4_{(10)}[C(i)S(R)+2 C(i)^2-3 C(G)C(i)] \eea Again
they are field wise and independent component wise diagonal. Only
the first term requires attention. The intermediate sums over
$n,r$ can be broken field wise and thereafter using diagonality of
the one loop anomalous dimensions  the first term    collapses to
a sum over intermediate connected irreps  weighted by their one
loop $\bar\gamma$ s: Thus for example \be
Y_{imn}Y^{npq}Y_{pqr}Y^{mrj}=Y_{im n_H}\bar\gamma^{(1)}_H Y^{m n_H
j}+Y_{i m n_{\Theta}} \bar\gamma^{(1)}_{\Theta} Y^{m
n_{\Theta}j}+... \ee  Thus the total contribution can be written
with the help of one loop anomalous dimension parameters. For
example :\bea \gamma_\Phi^{(2)}&=&-(240 |\eta|^2
(\bar{\gamma}_\Sigma^{(1)}+ \bar{\gamma}_{\bar{\Sigma}}^{(1)}  )+4
|k|^2(\bar{\gamma}_H^{(1)}+ \bar{\gamma}_\Theta^{(1)} )+6
|\gamma|^2(\bar{\gamma}_H^{(1)}+ \bar{\gamma}_\Sigma^{(1)} )
\nonumber\\&&+360|\lambda|^2\bar{\gamma}_\Phi^{(1)} +
4|\rho|^2\bar{\gamma}_\Theta^{(1)} +6
|\bar{\gamma}|^2(\bar{\gamma}_H^{(1)}+
\bar{\gamma}_{\bar\Sigma}^{(1)} )+60
|\zeta|^2(\bar{\gamma}_\Theta^{(1)}+ \bar{\gamma}_\Sigma^{(1)}
)\nonumber\\&&+60 |\bar{\zeta}|^2(\bar{\gamma}_\Theta^{(1)}+
\bar{\gamma}_{\bar\Sigma}^{(1)} ))+g^2_{10}(6240|\eta|^2+ 24|k|^2
+4320|\lambda|^2+36|\rho|^2\nonumber\\&&+60|\gamma|^2+60|\bar\gamma|^2+1320
|\zeta|^2+1320 |\bar{\zeta}|^2)+3864 g^4_{10}\eea

 \section{Probing the deep cleft : Applications of NMSGUT RG equations }

\subsection{ Landau Polar versus Emergence domain}
Let us begin with the elephants in the room : the huge $\beta$
function coefficients  in  the 1-loop gauge $\beta$-function and
also in  the $\beta$ functions of  almost all the chiral multiplet
self couplings in the superpotential. Thus the coefficients of the
cubic terms for the couplings
$\{g_{10},\lambda,\eta,\gamma,\bar\gamma
,\kappa,\zeta,\bar\zeta,\rho\}$   are seen  from
eqns.(\ref{betag1loop})-(\ref{gammaPhi1})  and the Appendix to be
$(16 \pi^2)^{-1}$ times   $\{137,180,640,142,142,95,265,265,16\}$
! Except for the couplings $\kappa,\rho$ the other couplings grow
even faster than the gauge coupling. As noted\cite{trmin,tas}
before the huge gauge $\beta$ functions   imply very rapid change
of $g_{10}$ and lead to a Landau pole in  the gauge coupling at
scales within an order of magnitude or so of the perturbative
unification scale. For the  usual  (SU(5) normalization) value of
the gauge coupling at unification :
$\alpha_5^{-1}(M_X^0)=\alpha_{10}^{-1}(M_X^0)/2=25.6$ we find the
SO(10) gauge coupling   has a Landau pole at about
$\Lambda_E\simeq exp(4\pi /137 \alpha_5(M_X^0))\simeq  10.5
M_X^0$. In the NMSGUT, even with the multitude of threshold
corrections,     $\alpha_5^{-1}(M_X^0)$   can consistently lie in
(at most)  the range $10-40$. This corresponds to $\theta_X \equiv
Log_{10}(\Lambda_E/M_X)$ varying in the range $\theta_X
\in[0.4,1.6]$ although the extreme values  are hard to achieve.
Thus the furthest that one can push the Landau Polar boundary  i.e
the scale beyond which the theory is certainly fully strong
coupled is about $10^{17.4}$ GeV.  Note however that this `UV
misbehaviour' pales in comparison with the effect of the combined
growth of the Yukawa couplings which can reach strong coupling
over an scale change by 20\% or less ! In fact we find that this
is true of   fits found by us earlier when they are extrapolated
into the UV. The strong divergence and instability of the
trans-unification flow into the UV implies that  it is not
efficient to look for fits to the complete SM  data by parameters
thrown at $M_X$ if one wishes to have any significant range of
energies where the SO(10) GUT is perturbative and well defined. It
is quite likely that the parameters optimized for such a  fit will
prove to lead to a Landau pole just above $M_X$.
Thus, to resolve this numerical difficulty,  we propose to turn the strong \emph{decrease in
couplings in the flow into the IR} to good account by searching
for viable coupling flows valid in the entire range $[M_X,
M_E\simeq \Lambda_E]$ of definition  by throwing the parameters
-subject to perturbative consistency constraints- at a candidate
$M_E$  and considering \emph{downward} (i.e. into the IR) running
of couplings. The scale $M_E$ ($M_X^0 < M_E << M_{Planck}$)   is
defined to be the scale where a (weakly coupled) effective GUT
with soft Susy breaking has emerged as supernatant to the unknown
strongly coupled dynamics of
  trans-emergence scales lying in $(\Lambda_E,M_{Planck})$.
  The strong RG flows, as well as the thousands of superheavy
  particles in the theory with masses $\sim M_X$   make a value of $M_E $ well
  below $ M_{Planck}$ plausible without forcing it
to coincide with the usual unification scale $M_X$. At least
\emph{prima facie}, $M_E$ could lie anywhere up to $10^{17.4}$
GeV. We accept that couplings will enter the Landau Polar region
at $\Lambda_E$ just above $M_E$ with the reassurance that by
choosing the initial values at $M_E$ the strongly weakening effect
of flow to lower scales will reduce the couplings further and
ensure that the theory becomes more accurately weakly coupled as
it approaches the region where the GUT crosses over into the low
energy effective theory i.e. the MSSM (with seesaw suppressed
neutrino mass and other GUT scale suppressed exotic operator
supplements). This pattern of energy scales is consistent with the
expectation \cite{gia} that the large number of massive degrees of
freedom in the NMSGUT will lead to the   the Planck mass being
dominated by their contribution to graviton wave function
renormalization : cutoff by the scale of the NMSGUT Landau Poles.
The Landau Polar latitude $\Lambda_E$ is set by examining the UV
flow from the values found to define an consistent low energy
theory and essentially coincides with $M_E$ : it is to be
interpreted as a perturbative limit and thus physical cutoff of
the effective SO(10) GUT signalled by the theory itself. It also
marks the point where a peculiar and mysterious  condensation
analogous to confinement in QCD but arising from UV flows takes
place in the SO(10) gauge dynamics. In sum, to probe viable
scenarios we should take $M_E$ to be   free along with the hard
and soft parameters defined at $M_E$ and conduct searches by first
running these parameters to a matching scale $M_X$ close to the
standard MSSM unification scale $M_X^0\simeq 10^{16.33}$ GeV and
then -after applying threshold corrections at that
scale\cite{bstabhedge} run the resultant effective MSSM parameters
down to the Electro-weak scale $M_Z$ and there -after applying low
scale threshold corrections- match them to the observed standard
model values.  This will be attempted in future improvements of
our fitting code. Here however we content ourselves by showing
that even running couplings down over the short interval between
the soft Susy parameter emergence scale $M_E$  and the GUT
matching scale can radically reshape the Susy breaking parameter
spectrum and bring it closer to the type of parameter values we
assumed in earlier studies\cite{ag2,nmsgut}.

\subsection{NMSGUT Running down  }
To illustrate  the actual  effect of running down the couplings
using the 1-loop $\beta$ functions, augmented by two loop results
for the gauge coupling and gaugino masses, we present an example
of a flow down from $M_E=10^{17.4}$ GeV where  $g_{10}(M_E)=1.0$
is quite small enough so that RGE flows are perturbative  yet
large enough that $g_{10}(M_X) \simeq 0.5$ required to match the
MSSM unification value without large threshold corrections can be
achieved. We note that   beginning with $g_{10}$ near to $3.0$ at
the Planck scale(roughly where the RG equation solution by 4th
order Runge-Kutta methods becomes unstable)  one can still flow
down close to unification gauge couplings $g_{10} \sim 0.5 $ such
as those found for the MSSM unification coupling.  Thus the strong SO(10)
gauge RG flows  into the infrared, as well as the GUT induced
gravity scenario, provide a rationale for effective separation of
the inevitable gauge-Yukawa  condensed strong coupling region in
the UV from the weakly coupled Susy SO(10)  GUT compatible with
the gauge and fermion data. These strong flows
can explain and justify certain features of the parameter values
assumed  in the extensive studies we have performed elsewhere\cite{nmsgut,bstabhedge} 
to find fits of the standard model parameters by matching to the effective MSSM obtained 
from the GUT after RG flow from $M_X$ to $M_Z$.  

 In our   example  we  take SO(10) gauge and Yukawa
couplings  similar to those found in earlier NMSGUT fits\cite{nmsgut,bstabhedge}. Examples of these features from the two explicit fits found in\cite{bstabhedge} are quoted in Table II. We see that these parameter choices exhibit the following features :\begin{itemize} \item Small    values of the gaugino masses $m_{1/2} << M_{susy}$ qualifying them to be considered as an induced as a secondary effect of the scalar soft Susy breaking parameters $m_0,A_0$ which are in the multiTeV to 100 TeV range \item Large negative values $\sim - (100 \,TeV)^2 $ for  the soft breaking parameters associated with the light Higgs doublets: i.e  mass squared values $M^2_{H,\bar H}$ and B parameter (soft analog of $\mu$ parameter in the superpotential)   \item $\mu$ parameter for the light Higgs doublets in 100 TeV range.\end{itemize}
\begin{table}
 $$
 \begin{array}{|c|c|c|c|}
 \hline 
 \mbox{Solution 1 Soft parameters}&{\rm m_{\frac{1}{2}}}=
          -152.899
 &{\rm m_{0}}=
         11400.993
 &{\rm A_{0}}=
         -2.0029 \times 10^{   5}
 \\
 \mbox{at $M_{X}$}&\mu=
          1.5966 \times 10^{   5}
 &{\rm B}=
         -1.7371 \times 10^{  10}
  &{\rm tan{\beta}}=           51.0000\\
 &{\rm M^2_{\bar H}}=
         -2.0655 \times 10^{  10}
 &{\rm M^2_{  H} }=
         -1.7978 \times 10^{  10}
 &
  \\
 \hline\hline
 \mbox{Solution 2 Soft parameters}&{\rm m_{\frac{1}{2}}}=
             0.000
 &{\rm m_{0}}=
         12860.405
 &{\rm A_{0}}=
         -1.9844 \times 10^{   5}
 \\
 \mbox{at $M_{X}$}&\mu=
          1.7240 \times 10^{   5}
 &{\rm B}=
         -1.4927 \times 10^{  10}
  &{\rm tan{\beta}}=           50.0000\\
 &{\rm M^2_{\bar H}}=
         -2.9608 \times 10^{  10}
 &{\rm M^2_{  H} }=
         -2.8920 \times 10^{  10}
 &
  \\
 \hline\hline
 
 \hline\end{array}
 $$
  \caption{\small{Examples of  EWSB relevant parameters  and  soft Susy breaking  parameters at $M_X$ from
  explicit fits with GUT scale threshold corrections \cite{bstabhedge} }}\end{table}
   
Searches for parameters using combined trans and cis unification RG flows are not attempted here. We restrain
our numerical investigations to showing that soft breaking parameters like those seen in Table II can be generated
by choosing soft breaking parameters according to canonical kinetic term SUGRY form
: all gaugino masses are zero, all soft scalar masses equal the
gravitino mass at the UV emergence scale: $m_{scalar}(M_E)=
m_{3/2}(M_E)$ and (for illustration)  $A_0(M_E)$=2$m_{3/2}$. We
also require that the soft bilinears obey the strictest form of
the gravity mediated scenario\cite{weinberg}: \be b_i =( A_0 -
m_{3/2} ) \mu_i\ee at the emergence  scale. We chose
$m_{3/2}(M_E)$=5 TeV and examine the renormalization flow  from
$M_{E}$ to $M_X^0$. The values of hard and soft parameters at two
scales ($M_E$ and $M_X^0$) are given in Tables \ref{t1} and
\ref{t2}. Clearly  the RG evolution can be very significant and in
particular the gauge coupling and soft masses change rapidly.
   Evolution of the Hermitian soft masses from $M_E$ to $M_X$  is
shown in Fig. \ref{softmassplot} and we can see that some of them
become negative. Moreover some of the $B$ parameters also turn
negative. In our realistic fits\cite{nmsgut,bstabhedge} we in fact
find that the values of soft hermitian masses squared and $B$
parameter relevant to the light MSSM Higgs at the GUT scales need
to be negative(Table II)  : which at least the  cSUGRY framework, applied directly at $M_X$,  would
contraindicate. The strong RG flows show that this constraint can
 be easily evaded if $M_E$ and $M_X$ do not coincide.

 $H,\bar H$  are constrained to be very light compared to the GUT scale
   by imposing $det{ \cal{H}}=0$ on their  mass matrix (${\cal{H}}$)
   which is calculated  using the MSGUT vevs\cite{abmsv,bmsv,fuku,ag1,ag2,nmsgut}. The
left and right null eigenvectors of ${\cal{H}}$ furnish the
``Higgs Fractions'' \cite{abmsv,ag2,nmsgut} whereby the
composition of the light doublets in terms of 6 pairs of GUT
doublets is specified and the rule for passing to the effective
theory : $ h_i \rightarrow
 \alpha_i  H, \bar{h}_i \rightarrow
 \bar{\alpha}_i  \bar{H}  $ defined. Then the soft
hermitian scalar mass terms will give \bea && m_i^2 (h_i^\dagger
h_i + \bar h_i^\dagger \bar h_i) \rightarrow m_H^2 H^\dagger H +
m_{\bar H}^2 \bar H^\dagger \bar H\nnu &&   m_H^2  = \sum_{i=1}^6
|\alpha_i|^2{m}_i^2 \quad ; \quad m_{\bar H}^2 = \sum_i^6 |\bar
\alpha_i|^2 {{\bar m}}_i^2\eea Since $m_i^2$ can turn negative
when running from $M_E$ to $M_X$ we see that negative
$m_{H,\overline H}^2$  can be achieved. However note that one also
has the $b_{ij}$ terms for each of the GUT Higgs multiplets so
that one will in fact also induce the B term for the light Higgs
as \bea B= b_{H} \alpha_1 \bar\alpha_1 +b_{\Sigma } (\alpha_3
\bar\alpha_2 +\alpha_2 \bar\alpha_3) +b_{\Phi} \alpha_4
\bar\alpha_4 +b_{\Theta } (\alpha_5 \bar\alpha_5 + \alpha_6
\bar\alpha_6)   \eea An additional constraint(analogous to that
imposed on $\mu$) to maintain the B term at magnitudes less than
$10^{10}$GeV$^2$ (rather than the RG evolved values which tend to
have magnitude $ (A_0-m_{3/2}) \mu_i \sim M_X m_{3/2}
>> m_{3/2}^2 $) is thus required. The turning of sign of
some of the $b$ parameters may  provide  a mechanism whereby the
short flow lands these parameters closer to the TeV scale values
required.
\begin{table}
 $$
 \begin{array}{ccc}
 \hline \hline&&\vspace{-.3 cm}\\ {\rm  Parameter }&
 {\mbox {Value at $M_E=10^{17.4}$ GeV} }& {\mbox {Value at $ M_{X}^0(10^{16.33}$ GeV}) }\vspace{0.1 cm} \\
 \hline\hline
 g_{10},g_5& 1.0, {\sqrt{2}} & 0.497,0.703\\

\lambda,\eta &  -0.0434 + 0.0078 i,-0.313 +0.08i & -0.0133 + 0.0024 i,-0.121 + 0.031 i  \\


\rho,\kappa &   0.954  - 0.27  i,0.027  + 0.1 i&0.21 - 0.06 i,0.0024  + 0.0088 i \\

\gamma,\bar\gamma  &   0.471 ,-3.272& 0.0493 ,-0.425  \\

 \zeta,\bar \zeta&   1.009  +  0.831 i,0.36 + 0.59 i & 0.265  + 0.218  i,  0.117  + 0.192  i \\

\hline\hline
   h_{11}/10^{-6} & 4.4602   & 1.241  \\
h_{22}/10^{-4} &  4.1031  & 1.1411  \\
 h_{33} &  0.0244  &  .00679 \\
 h_{12}/10^{-12} & 0.0  & -1.816+2.919i  \\
h_{13}/10^{-11} &  0.0  & -1.823+1.811i  \\
 h_{23}/10^{-9} &  0.0  &  -2.955+5.549i \\
     f_{11}/10^{-6} &  -.0044+.16207 & -0.0045+.166 i\\
f_{22}/10^{-5} & 6.675+4.8457i   & 6.843+ 4.968i\\
 f_{33}/10^{-4} & -9.264+2.7876i   &  -9.498+2.858i \\
 f_{12}/10^{-6} & -0.849-1.782 i  &  -.871-1.828i  \\
f_{13} /10^{-6}&  .5496+1.1479 i  &  0.5635+1.177i \\
 f_{23} /10^{-4} &  -.4266+2.231i  &  -0.4374+2.287i \\
 g_{12}/10^{-5} &  1.4552+1.599i  &  1.016+ 1.116 i \\
g_{13}/10^{-5} & -11.784+4.9613i   & -8.227+3.464i  \\
 g_{23}/10^{-4} &  -1.6648-1.18436i  &  -1.162-0.827i \\
     \hline\hline
     \mu_\Phi & 10^{15}~\mbox{GeV}& 4.55 \times 10^{14}~\mbox{GeV}  \\
    \mu_H &10^{15} ~\mbox{GeV}&  5.23 \times 10^{13}~\mbox{GeV} \\
\mu_\Sigma &  10^{15} ~\mbox{GeV}  & 5.72 \times 10^{14}~\mbox{GeV}  \\
  \mu_{\Theta} &  10^{15}~\mbox{GeV} & 3.29 \times 10^{14}~\mbox{GeV} \\
   \hline
 \end{array}
 $$
\caption{\small{Example of  consistent  hard  NMSGUT-cSUGRY
parameters emergent at   $M_E=10^{17.4}$ GeV evolved down to
$M_X^0=10^{16.33}$ GeV   using one-loop NMSGUT RGEs for all
parameters except the gauge coupling and gaugino mass which use
two loop evolution. \label{t1}}}
 \end{table}

\begin{figure}[tbh]
\centering
\includegraphics[scale=1.6]{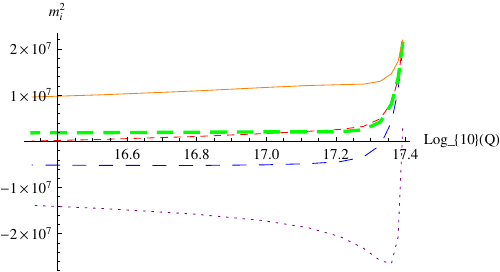}
\caption{Evolution of soft masses from Planck scale to GUT scale.
Dashed (red), dotted (purple), medium dashed (blue), thick dashed
(green) and solid (orange) lines represent ${m}^2_{\bar{\Phi}}$,
${m}^2_{H}$, ${m}^2_{\Theta}$, ${m}^2_{\Sigma}$ and
${m}^2_{\bar\Sigma}$
respectively.\label{softmassplot}}\vspace{.5cm}
\end{figure}

\begin{table}
 $$
 \begin{array}{ccc}
 \hline\hline &&\vspace{-.3 cm}\\
 {\rm  Parameter }&\mbox {Value at $M_E=10^{17.4}$ GeV}& \mbox {Value at $ M_{X}^0(10^{16.33})$ GeV}\vspace{0.1 cm} \\
\hline\hline&&\vspace{-.4 cm}\\
 \tilde\lambda,\tilde\eta &-434.0 + 78.0 i,-3127.0 +798.0 i &   -17.47 + 3.14 i, -335.8 + 85.69 i\\
\tilde\rho,\tilde{ k} & 954.4 -269.8  i,273.0 + 991 i &-115.7 + 32.7 i,-6.39 -23.19 i \\
   \tilde\gamma,\tilde{\bar\gamma}  & 4711 +0.0i,-32719 +0.0 i   &-80.3 +0.116 i,142.7 +0.021i  \\
    \tilde{\zeta},\tilde{\bar \zeta}& 10091 + 8305  i,3596+ 5898 i & 125.28 + 103.1i,206.77 + 339.14 i  \\
     \hline  \hline&&\vspace{-.4 cm}\\

    \tilde h_{11}/10^{-4}  & 446.02   & 63.05+0.0028 i  \\
 \tilde h_{22},\tilde h_{33} &  4.10,244.19  &  0.58 +2.647 \times 10^{-5}i,34.52 + 0.00158 i \\
     \tilde h_{12}/10^{-8},\tilde h_{13}/10^{-7} & 0.0,0.0   & -3.65+ 5.88 i,-3.071+4.62 i  \\
 \tilde h_{23}/10^{-5} &   0.0   &  -7.072+ 13.2 i   \\
     \tilde f_{11}/10^{-3},\tilde  f_{22} & -0.0436+1.621 i,.667+0.4845i &-0.042+1.58i, 0.65+0.472 i \\
     \tilde   f_{33},\tilde   f_{12}/10^{-2} & -9.264+2.787 i,-0.85-1.78 i &-9.013+2.71i,-0.83-1.73i \\
    \tilde  f_{13}/10^{-2},\tilde   f_{23} & 0.55+1.15 i,-0.427+2.23 i & 0.535+1.12 i,-0.415+2.17 i\\
     \tilde  g_{12} & 0.146+0.16i & 0.073+0.08i \\
    \tilde  g_{13},\tilde  g_{23} & -1.178+0.496 i, -1.665 - 1.184 i  & -0.591+0.249 i,-0.835- .594 i \\
   \hline   \hline&&\vspace{-.7 cm}\\
     {M_{\tilde g}} & 0 & -1171.73 + 0.0016i \\\hline\hline&&\vspace{-.7 cm}\\
    {b_\Phi}  &  5.0 \times 10^{18} \mbox{GeV}^2 &  -3.605 \times 10^{17} + 6.576 \times 10^{12}i \mbox{GeV}^2  \\
 {b}_H & 5.0 \times 10^{18} \mbox{GeV}^2 & -3.579  \times 10^{17} + 2.474 \times 10^{13} i \mbox{GeV}^2  \\
{b}_\Sigma  & 5.0 \times 10^{18} \mbox{GeV}^2 & 3.881 \times 10^{17} + 6.82 \times 10^{12} i\mbox{GeV}^2  \\
 {b}_{\Theta} & 5.0 \times 10^{18} \mbox{GeV}^2 & -8.72  \times 10^{17}- 8.536 \times 10^{11} i \mbox{GeV}^2 \\
 \hline  \hline&&\vspace{-.4 cm}\\
  {m}^2_{\Phi} &  2.5 \times 10^7 \mbox{GeV}^2 &48070.7 \mbox{GeV}^2  \\
 {m}^2_{H} & 2.5 \times 10^7 \mbox{GeV}^2  & -1.388 \times 10^7 \mbox{GeV}^2 \\
   {m}^2_{\Theta}& 2.5 \times 10^7 \mbox{GeV}^2 &-5.154 \times 10^{6} \mbox{GeV}^2 \\
{m}^2_{\Sigma}&  2.5 \times 10^7 \mbox{GeV}^2 & 1.80955 \times 10^6 \mbox{GeV}^2\\
   {m}^2_{\bar{\Sigma}}&  2.5 \times 10^7 \mbox{GeV}^2  &9.564 \times 10^6  \mbox{GeV}^2 \vspace{0.1 cm}\\
    {\rm Eval~{m}^2_{\tilde\Psi}}& 2.5 \times 10^7 \mbox{GeV}^2 &\{2.7892,2.7892,2.7889\} \times 10^7 \mbox{GeV}^2 \vspace{0.1 cm}\\
     \hline\hline
 \end{array}
 $$
 \label{table c}
 \caption{\small{Values of NMSGUT soft parameters at two different scales evolved by using one-loop
  SO(10) RGEs.}   $A_0=10$ TeV, $m_{3/2}=5$ TeV.\label{t2}}

 \end{table}
 \vspace{.5cm}

\section{Concerning Invariant parameter  submanifolds under the  MSGUT RG flow  }

The seminal work of Pendleton and Ross\cite{pendlross} on quasi
fixed points of the SM RG flow  successfully estimated the
approximate top quark mass before its discovery on the basis of
the intuition of an approximate ``quasi infra red  fixed point''
in the RG flow governing the ratio $h_t^2/\alpha_s$. Since then
the same basic idea has been applied to the dimensionless and even
dimensionful(i.e. soft)  parameters of (Susy) GUTs to study
 \cite{lanzagross} whether the quasi-fixed point structure of the GUT Yukawa
and gauge couplings may be significant in fixing the couplings at
the Unification scale. It was found that such structures are
particularly relevant in the case where there are many fields so
that the GUT model is strongly coupled in the ultraviolet. This is
precisely the case for the MSGUT and NMSGUT.  If such invariant
structures could be identified they would obviously be an
important criterion for comparing different unified models. In the
present instance we have calculated the \emph{full} set of RG
equations for both hard and soft couplings of the  MSGUT. Thus the
optimistic view might be that these complex flow equations somehow
support novel invariant structures when considered in their
entirety. The generic form of the 1 loop
beta functions for the dimensionless (gauge and Yukawa ) couplings
of a supersymmetric model is \bea \beta_g=&=& {\frac{1}{16 \pi^2}}
b_0 g^3  \nnu \beta_{Y^{i j k}} &=& {\frac{1}{16 \pi^2}} (
\gamma_i + \gamma_j +\gamma_k) Y^{ijk}\eea  where $\gamma_i$ is given by
eqn(\ref{gammai}) after using   diagonality ( $\gamma^i_j\equiv \gamma_i \delta^i_j$) of the anomalous dimension matrices ,   and $b_0$ is a large integer or rational number
($b_0=137$ for the MSGUT ). It is clear that combining these
two equations and the 1-loop formula for $\gamma_i$ we can derive
fixed point conditions in terms of $Z_{ijk}=|Y_{ijk}|^2/g^2$ for
the squared magnitudes of  various couplings , while their phases
remain free. These conditions are readily seen to be generically
of the form \bea \gamma_i + \gamma_j +\gamma_k  - b_0 g^2
=0={\frac{ \bar{\gamma}_i + \bar{\gamma}_j +\bar{\gamma}_k }{g^2}}
-(b_{ijk} +b_0)  \eea

where we have separated out the gauge ($b_{ijk}$)and Yukawa($\bar{\gamma}_{i,j,k}$) components of the
anomalous dimensions for fields i,j,k. Writing $\bar\gamma_i=
a_i^I |Y_I|^2 $ where $I$ runs over the different Yukawa couplings
in the theory we get the fixed point conditions in the form \bea
(a_i^I + a_j^I  + a_k^I ) Z_I -(b_{ijk} +b_0) =0 \eea where
$Z_I=|Y_I|^2/g^2$.

The question as to whether any quasi fixed points of the full set
of RG equations can possibly exist  then involves solving these
equations subject to the constraints that all $Z_I$  are positive
semi-definite.  Unfortunately the huge value of the coefficient
$b_0$ which is common to all the conditions makes a solution
impossible to achieve.

We illustrate the difficulty for a simplified MSGUT
model with   negligible first generation matter Yukawas
$(h,f,g)_{1A}\simeq 0$,diagonal $h,f$ couplings $h_{2,3},f_{2,3}$ and
$g_{32}=-g_{23}$. 
The relevant anomalous dimensions are \bea
{\frac{\bar\gamma_\Phi}{g^2}} &=& 4 Z_k + 180 Z_\lambda + 2 Z_\rho +
240 Z_\eta + 6(Z_\gamma + Z_{\bar\gamma}) + 60 (Z_{\zeta}
+Z_{\bar\zeta}) \nnu {\frac{\bar\gamma_{\bar\Sigma}} {g^2}} &=&    
200 Z_\eta + 10 Z_{\bar\gamma} + 100 Z_{\bar\zeta}  + 32(Z_{f_2}
+Z_{f_3}) \nnu {\frac{\bar\gamma_\Sigma}{g^2}}&=&  200 Z_\eta +
10 Z_{\gamma} + 100 Z_{ \zeta}  \nnu {\frac{\bar\gamma_H}{g^2}}&=&
84 Z_k +   126(Z_\gamma + Z_{\bar\gamma}) + 8(Z_{h_2} +Z_{h_3})
\nnu {\frac{\bar\gamma_\Theta}{g^2}}&=&  7 Z_k +7 Z_\rho + 105
(Z_{\zeta} +Z_{\bar\zeta}) + 16 Z_{g_{23}} \nnu
{\frac{\bar\gamma_\psi^T}{g^2}}&=& 252 (0,Z_{f_2},Z_{f_3}) + 120
(0,Z_{g_{23}},Z_{g_{23}}) + 10(0,Z_{h_2},Z_{h_3})\eea

Consider first the case where the couplings to the 16-plets have been set to zero($Z_{h_{2,3},f_{2,3},g_{23}}\equiv 0$). The fixed point conditions for the other couplings are then 
 \bea
 0 &=&
3 {\frac{\bar\gamma_\Phi}{g^2}} - (b_0 + 72)  \nnu 0 &=&
{\frac{(\bar\gamma_\Phi + \bar\gamma_\Sigma  +\bar\gamma_H)}{g^2}} - (b_0 + 58)  \nnu
   0 &=&
{\frac{(\bar\gamma_\Phi +{ \bar\gamma}_{\bar\Sigma} + {\bar\gamma_H})}{g^2}}  - (b_0 + 58) \nnu
  0 &=&
{\frac{(\bar\gamma_\Phi + \bar\gamma_\Sigma +\bar\gamma_{\bar\Sigma} )}{g^2}} - (b_0 + 74) \nnu 0 &=&
{\frac{(\bar\gamma_\Theta + \bar\gamma_H +\bar\gamma_\Phi )}{g^2}}
- (b_0 + 54) \nnu 0 &=& {\frac{(2\bar\gamma_\Theta
+\bar\gamma_\Phi )}{g^2}} - (b_0 + 66) \nnu
 0 &=& {\frac{(\bar\gamma_\Phi +
\bar\gamma_\Sigma +\bar\gamma_{\Theta} )}{g^2}} - (b_0 + 70) \nnu 0
&=& {\frac{(\bar\gamma_\Phi + \bar\gamma_\Theta
+\bar\gamma_{\bar\Sigma} )}{g^2}} - (b_0 + 70) \eea

Solving these fixes $Z_{\eta,k,\zeta,\bar\zeta,\rho}$ in terms of  $Z_{\lambda,\gamma,\bar\gamma}$ and for them to be semipositive  gives 5 inequalities which can be easily reduced by eliminating $Z_{\gamma}$ between them. However this yields the condition $  Z_{\lambda}\leq  -227/2700$ which is   inconsistent with the semipositive values allowed for $Z_{\lambda}$. So there is no fixed point. 

One might hope that introducing the 16-plet couplings might help. Then we restore $Z_{h_{2,3},f_{2,3},g_{23} }$ and obtain the additional conditions for the  beta functions of these ratios : \bea    0 &=&
{\frac{(\bar\gamma_{H} + 2 \bar\gamma_{\psi_{2,3}})}{g^2}} - (b_0 + 63/2) \nnu0 &=&
{\frac{(\bar\gamma_{\bar\Sigma} + 2 \bar\gamma_{\psi_{2,3}})}{g^2}} - (b_0 + 95/2) \nnu 0 &=&
{\frac{(\bar\gamma_\Theta +   \bar\gamma_{\psi_2}
+\bar\gamma_{\psi_3} )}{g^2}} - (b_0 + 87/2) \eea

Solving these conditions one finds that
$Z_{\zeta,\bar\zeta,h_3,f_2,f_3,g_{23},\rho}$ are determined in
terms of $Z_{\eta,\gamma,\bar\gamma,k,\lambda,h_2}\geq 0$ which are
themselves undetermined. The question is whether there are any
semipositive  values of these free parameters for which the
dependent variables remain semipositive.  Solution of the fixed point conditions 
  yields the following solution vector  
    \bea && \{Z_\zeta,Z_{\bar\zeta},Z_{h_3},Z_{f_2},Z_{f_3},Z_{g_{23}},Z_\rho \}=\nnu &&  \{53/75 - 2
Z_\eta - Z_\gamma /10,689/525 - 10 Z_\eta  -  
Z_{\bar\gamma}/10   - 12 Z_\lambda,\nnu &&
  41/6 - (63 Z_\gamma)/4 - (63 Z_{\bar\gamma})/4 - Z_{h_2} - (21 Z_k)/
  2, \nnu && -(613/756) + (25 Z_\eta)/2 - (5Z_\gamma)/16 - (5 Z_{\bar\gamma})/16 - (
  5 Z_{h_2})/126 -5 Z_k/24 + (75 Z_\lambda)/4,\nnu &&
   -(409/378) + (25 Z_\eta)/2 + (5 Z_\gamma)/
  16 + (5 Z_{\bar\gamma})/16 + (5Z_{h_2})/126 + (5 Z_k)/24 + (75 Z_\lambda)/4,\nnu &&
 209/96 - (105 Z_\eta)/4 + (21 Z_\gamma)/32 + (21 Z_{\bar\gamma})/32 + (
  7 Z_k)/16 - (315 Z_\lambda)/8,\nnu &&
   -(1081/42) + 240 Z_\eta   - (2 Z_k)  + 270 Z_\lambda  \} \geq 0
\eea
An elementary reduction of this system of inequalities \cite{solod}   leads to contradictory condition  
\bea  -(407+252 Z_k+11340 Z_\lambda)/48 \geq Z_{h_2}  \geq 0\eea showing that again  there is no fixed point for the
system even when  measuring in units of the (exploding in the UV) value of $g^2$.  Although we have not obtained
a general proof it seems likely that no fixed point can be found.  Support for this can be found in  recent investigations \cite{bajcsann}, based
on the so called non-perturbative a-theorem  and the exact NSVZ beta function(see \cite{bajcsann}  for a concise
introduction and a fairly complete list of references for these topics), of the possibility of non-trivial superconformal
UV fixed points in the SO(10) MSGUT.   They  conclude that no such fixed points exist  without rather artificial
requirements being placed upon the couplings and R-charges of some of the SO(10) multiplets or by introducing
very large numbers of additional multiplets and trivializing the superpotentials allowed.

\section{Discussion}
We have proposed a  framework for a consistent interpretation of
Asymptotically Strong GUTs by considering RG flows of GUT
parameters from an emergence scale $M_E$ of a weakly coupled GUT
down to the scale $M_X$  where the GUT is matched to its low
energy effective theory.  Thereafter the MSSM flows  from $M_X$
down to $M_Z$ determine the experimental predictions of the GUT
parameter set chosen at $M_E$. This procedure allows extension of
the perturbative regime of the unified theory up to  the Landau
Polar latitude $\Lambda_E$.   Interestingly the large number of
degrees of freedom further strengthen the
intuition\cite{trmin,tas} that the scale of Gravity may be
dominantly set by the effects of (the thousands of)  NMSGUT
superheavy particles. Thus $M_{Pl}$ can lie well above $\Lambda_E$
which nevertheless plays a part in raising $M_{Pl}$ by serving as
the physical  cutoff scale for graviton wave function
renormalization corrections due to the NMSGUT as well as the scale
for SO(10) ``confinement'' \cite{tas}. We presented the NMSGUT RG
equations to determine the RG evolution of couplings between the
scale($M_E$) where  the perturbative effective theory (NMSGUT plus
weakly coupled and softly broken $N=1$ supergravity) emerges and
the matching scale between GUT and the low energy effective theory
(i.e the MSSM)  at $M_X$. To illustrate the application of these
results we evaluated the effects of the 1-loop evolution on
randomly chosen sets of parameter values assuming a minimal,
canonical kinetic term, supergravity scenario for the starting
parameter ansatz. From the Tables and Fig. 1  we see that the RG
evolution has dramatic effects on the soft Susy breaking
parameters. Firstly most of the soft Susy breaking Hermitian
masses  squared of   the SO(10) Higgs irreps become negative even
though they start from a common positive mass. This provides a
potentially robust justification of the negative values of
$M_{H,\bar H}^2$  used in NMSGUT fits \cite{nmsgut,bstabhedge}.
Note that the distinctive normal s-hierarchy at low scale is
strongly correlated with the large negative $M_{H,\bar H}^2$ we
use in the fits. Gaugino masses($M_{\lambda}$)  will be generated
by two loop RG evolution between $M_X$ and $M_Z$,
  even if  $M_{\lambda}$=0 at the  scale $M_X$. On the other hand the same applies to the evolution
   between $M_E$ and $M_X$. Thus even canonical gauge kinetic terms in the GUT can still generate adequate
   gaugino masses. This is  pleasing since we have always resisted invoking non
  canonical  K$\ddot{a}$hler potential  and gauge kinetic terms on grounds of minimality/predictivity and to preserve
   renormalizability of the gauge sector.

     Another notable effect is the
intermediate scale $(O({m_{3/2}M_X}))$ values of the soft
 parameters $b_{\Phi,\Sigma,\Theta,H}$ required by the canonical mSUGRY
  ansatz and induced by the dependence $\frac{db}{dt}$ $\sim$ $M_X m_{3/2}$.
  So we may need to impose an additional condition
   in order  that the contribution from the soft terms to $b_{H,\bar H}$ is
    O($m_{3/2}^2$) unless this is  achievable via the RG flow of
    $B_{ij}$ towards negative values itself.

    The running of trilinear soft coupling and s-fermion mass squared parameters
     ($m_{\tilde\Psi}^2$) will  give distinct values
     at the GUT scale for the three generations  (considered the 
    same in earlier studies of     NMSGUT\cite{nmsgut,bstabhedge}).
        In   sequels we will integrate  these RG flows with our
    previous code that incorporates the MSSM flows between $M_X$
    and $M_Z$. Then one will throw the core soft parameters
     $m_{3/2},A_0$ at $M_E$ and run down over thresholds
    to $M_Z$ with one additional fine tuning constraint. Thus the total
    number of soft parameters will be significantly reduced. Improvements would include the 2-loop RG coefficients we have already computed \cite{phds}.
Finally the straightforward (since the superpotential vertex
connectivity is preserved)  generalization of these results to the
case of YUMGUTs\cite{yumgut} will allow us also to perform the RG
flows from the Planck scale for dynamical flavour generation
models based on the MSGUT. These theories have around 6 times as
many fields as the NMSGUT and are thus even more capable of
separating $M_X$ and $M_{Pl}$. We note that the techniques we have
used to actually
 evaluate the 2-loop RGEs have overcome the combinatorial
 complexity that prevented their calculation by automated means.
 They can be used for any Susy GUT.

\section{Acknowledgements}  CSA acknowledges financial support from the the  Department of Science and Technology,
Government of India , under SERB Project No. EMR/2014/000250  on the `` Phenomenology and Cosmology of the New Minimal Supersymmetric SO(10) GUT''.

\newpage
\section*{Appendix }
\subsection*{One-loop RGEs\label{oneloop}}
One-loop anomalous dimension parameters associated with different
superfields:

\bea \gamma_i^{(1)j}&=&\frac{1}{2} Y_{ipq}Y^{jpq}-2 g_{10}^2
\delta_i^j C(i) \nonumber \\ \gamma_\Sigma^{(1)}&=& 200 |\eta
|^2+10 |\gamma |^2+100 |\zeta |^2-25 g_{10}^2 \nonumber \\
\gamma_{\bar\Sigma}^{(1)}&= &200 |\eta |^2+10 |\bar{\gamma}|^2+100
|\bar{\zeta}|^2+32 \text{Tr}[f^{\dag}.f]-25 g_{10}^2 \nonumber
\\
 \gamma_H^{(1)}&=& 84 |\kappa |^2+126
(|\gamma|^2+|\bar{\gamma}|^2)+8 \text{Tr}[h^{\dag}.h] -9 g_{10}^2
\nonumber \\ \gamma_\Theta^{(1)}&=& 7 (|\kappa |^2+|\rho |^2)+105
(|\zeta |^2+|\bar{\zeta}|^2)+8 \text{Tr}[g^{\dag}.g]-21 g_{10}^2
\nonumber
\\
(\gamma_\Psi^{(1)})_A^B&=& (\gamma_\Psi^{(1)})_{AB} =252
f^{\dag}.f+120 g^{\dag}.g+10 h^{\dag}.h-\frac{45 g_{10}^2}{4} \eea

\bea \bar\gamma_i^{(1)j}&=&\frac{1}{2} Y_{ipq}Y^{jpq} \nonumber\\
\bar\gamma_\Sigma^{(1)}&=&  200 |\eta |^2 +10 |\gamma |^2+100
|\zeta |^2\nonumber\\
 \bar\gamma_{\bar\Sigma}^{(1)}&=& 200 |\eta |^2+ 10
|\bar\gamma |^2+100 |\bar \zeta |^2+32 \text{Tr}[f^{\dag }.f] \nonumber\\
\bar \gamma_H^{(1)}&=& 84 |\kappa |^2+126
(|\gamma|^2+|\bar{\gamma}|^2)+8 \text{Tr}[h^{\dag }.h] \nonumber\\
\bar\gamma_\Theta^{(1)}&=& 7(|\kappa|^2+|\rho |^2)+ 105
(|\zeta |^2+|\bar\zeta |^2)+8\text{Tr}[g^{\dag }.g]  \nonumber\\
 \bar\gamma_\Psi^{(1) }&=&252 f^{\dag}.f+120 g^{\dag}.g+10
h^{\dag}.h \eea

One-loop beta functions for the SO(10) superpotential parameters
and  Yukawa couplings are:

\be  \beta_\lambda^{(1)}=  3 \gamma_\Phi^{(1)}  \lambda  \quad ;
\quad \beta_\eta^{(1)}= \eta (\gamma_\Sigma^{(1)} +\gamma_{\bar
\Sigma}^{(1)} +\gamma_\Phi^{(1)} )   \label{one}     \ee

\be  \beta_\gamma^{(1)}= \gamma
(\gamma_H^{(1)}+\gamma_\Sigma^{(1)} +\gamma_\Phi^{(1)} )     \quad
; \quad  \beta_{\bar\gamma}^{(1)}= \bar\gamma
(\gamma_H^{(1)}+\gamma_{\bar\Sigma}^{(1)} +\gamma_\Phi^{(1)} ) \ee

\be  \beta_k^{(1)}= k (\gamma_H^{(1)}+\gamma_\Theta^{(1)}
+\gamma_\Phi^{(1)} )     \quad ; \quad  \beta_\zeta^{(1)}=
\zeta(\gamma_\Theta^{(1)} +\gamma_\Sigma^{(1)} +\gamma_\Phi^{(1)}
) \ee

\be  \beta_{\bar\zeta}^{(1)}= \bar\zeta(\gamma_\Theta^{(1)}
+\gamma_{\bar\Sigma}^{(1)} +\gamma_\Phi^{(1)} ) \quad ; \quad
\beta_\rho^{(1)}= \rho  (\gamma_\Phi^{(1)}+2 \gamma_\Theta^{(1)} )
\ee

\be  \beta_h^{(1)}= h
\gamma_H^{(1)}+(\gamma_\Psi^{(1)})^{T}.h+h.\gamma_\Psi^{(1)} \quad
; \quad  \beta_f^{(1)}= f
\gamma_{\bar\Sigma}^{(1)}+(\gamma_\Psi^{(1)})^{T}.f+f.\gamma_\Psi^{(1)}
\ee

\be  \beta_g^{(1)}= g
\gamma_\Theta^{(1)}-(\gamma_\Psi^{(1)})^{T}.g+g.\gamma_\Psi^{(1)}
\ee

\be \beta_{\mu_{\Phi}}^{(1)}=2  \gamma_{\Phi}^{(1)}\mu_{\Phi}
\quad ; \quad \beta_{\mu_H}^{(1)}=2 \gamma_{H}^{(1)} \mu_H\ee

\be
\beta_{\mu_{\Sigma}}^{(1)}=(\gamma_{\Sigma}^{(1)}+\gamma_{\bar{\Sigma}}^{(1)})
\mu_{\Sigma}\quad ; \quad  \beta_{\mu_{\Theta}}^{(1)}=2
\gamma_{\Theta}^{(1)} \mu_{\Theta} \label{last} \ee

\subsection*{Soft parameters RGEs\label{oneloopsoft}}

\bea \tilde\gamma_i^{(1)j}&=&\frac{1}{2} Y_{ipq}h^{jpq} \nonumber
\\ \tilde{ \gamma}_\Sigma^{(1)}&=& 200 \tilde{\eta}  \eta ^{*}+10
\tilde{\gamma} \gamma ^{*}+100 \tilde{\zeta} \zeta ^{*} \nonumber
\\  \tilde{ \gamma}_{\bar\Sigma}^{(1)}&=& 200 \tilde{\eta} \eta
^{*}+10 \tilde{\bar{\gamma}} \bar\gamma ^{*}+100 \tilde{\bar\zeta}
\bar\zeta ^{*}+32\text{Tr}[f^{\dag}.\tilde{f}]  \nonumber
\\   \tilde{ \gamma}_H^{(1)}&=& 84\tilde{\kappa} \kappa^{*} +
126(\tilde{\gamma} \gamma^{*} + \tilde{\bar{\gamma}}
\bar\gamma^{*}) +8\text{Tr}[h^{\dag}.\tilde{h}]\nonumber
\\ \tilde{ \gamma}_\Theta^{(1)}&=&7( \tilde{\kappa} \kappa^{*}+
\tilde{\rho}  \rho ^{*})+105 (\tilde{\zeta} \zeta
^{*}+{\tilde{\bar\zeta }} \bar\zeta^{*})+8
\text{Tr}[g^{\dag}.\tilde{g}] \nonumber
\\\tilde{ \gamma}_{\psi}^{(1)}&=&252 f^{\dag}. \tilde{f}+120
g^{\dag}.\tilde{g}+10 h^{\dag}.\tilde{h} \eea

\bea \hat\gamma_i^{(1)j}&=&\frac{1}{2} h_{ipq}h^{jpq} \nonumber
\\
\hat{\gamma}_{\Phi}^{(1)}&=&240|\tilde{\eta}|^2+4|\tilde{\kappa}|^2+180|\tilde{\lambda}|^2+2|\tilde{\rho}|^2+
6(|\tilde{\gamma}|^2+|\tilde{\bar{\gamma}}|^2)+60(|\tilde{\zeta}|^2+|\tilde{\bar{\zeta}}|^2)\nonumber
\\ \hat{\gamma}_{\Sigma}^{(1)}&=&200|\tilde{\eta}|^2+10
|\tilde{\gamma}|^2+ 100|\tilde{\zeta}|^2 \nonumber
\\ \hat{\gamma}_{\bar{\Sigma}}^{(1)}&=&200|\tilde{\eta}|^2+10
|\tilde{\bar{\gamma}}|^2+ 100|\tilde{\bar{\zeta}}|^2 +32 \text
Tr[\tilde{f}^{\dag}.\tilde{f}]\nonumber
\\ \hat{\gamma}_{H}^{(1)}&=&84|\tilde{\kappa}|^2+
126(|\tilde{\gamma}|^2+|\tilde{\bar{\gamma}}|^2)+8 \text
Tr[\tilde{h}^{\dag}.\tilde{h}]\nonumber
\\
\hat{\gamma}_{\Theta}^{(1)}&=&7(|\tilde{\kappa}|^2+|\tilde{\rho}|^2)+
105(|\tilde{\zeta}|^2+|\tilde{\bar{\zeta}}|^2)+8 \text
Tr[\tilde{g}^{\dag}.\tilde{g}]\nonumber
\\ \hat{\gamma}_{\Psi}^{(1)}&=&252
\tilde{f}^{\dag}.\tilde{f}+120\tilde{g}^{\dag}.\tilde{g}+10\tilde{h}^{\dag}.\tilde{h}
\eea

 One-loop beta functions for the soft
parameters:

\be \beta_{\tilde{\eta}}^{(1)}= \tilde{\eta}(
\bar{\gamma}_{\Phi}^{(1)}+ \bar{\gamma}_{\Sigma}^{(1)}+
\bar{\gamma}_{\bar{\Sigma}}^{(1)})+2\eta(
\tilde{\gamma}_{\Phi}^{(1)}+ \tilde{\gamma}_{\Sigma}^{(1)}+
\tilde{\gamma}_{\bar{\Sigma}}^{(1)}) -74g_{10}^2(\tilde{\eta}-2 M
\eta) \ee

\be \beta_{\tilde{\gamma}}^{(1)}= \tilde{\gamma}(
\bar{\gamma}_{\Phi}^{(1)}+ \bar{\gamma}_{\Sigma}^{(1)}+
\bar{\gamma}_{H}^{(1)})+2\gamma( \tilde{\gamma}_{\Phi}^{(1)}+
\tilde{\gamma}_{\Sigma}^{(1)}+ \tilde{\gamma}_{H}^{(1)}) -58
g_{10}^2(\tilde{\gamma}-2 M \gamma) \ee

\be \beta_{\tilde{\bar{\gamma}}}^{(1)}= \tilde{\bar{\gamma}}(
\bar{\gamma}_{\Phi}^{(1)}+ \bar{\gamma}_{\bar\Sigma}^{(1)}+
\bar{\gamma}_{H}^{(1)})+2\bar{\gamma}(
\tilde{\gamma}_{\Phi}^{(1)}+ \tilde{\gamma}_{\bar\Sigma}^{(1)}+
\tilde{\gamma}_{H}^{(1)}) -58g_{10}^2(\tilde{\bar{\gamma}}-2 M
\bar{\gamma}) \ee

\be \beta_{\tilde{\kappa}}^{(1)}= \tilde{\kappa}(
\bar{\gamma}_{\Phi}^{(1)}+ \bar{\gamma}_{\Theta}^{(1)}+
\bar{\gamma}_{H}^{(1)})+2\kappa ( \tilde{\gamma}_{\Phi}^{(1)}+
\tilde{\gamma}_{\Theta}^{(1)}+ \tilde{\gamma}_{H}^{(1)}) -54
g_{10}^2(\tilde{\kappa}-2 M \kappa) \ee

\be \beta_{\tilde{\rho}}^{(1)}= \tilde{\rho}(
\bar{\gamma}_{\Phi}^{(1)}+ 2 \bar{\gamma}_{\Theta}^{(1)})+2\rho(
\tilde{\gamma}_{\Phi}^{(1)}+ 2 \tilde{\gamma}_{\Theta}^{(1)}) -66
g_{10}^2(\tilde{\rho}-2 M \rho )\ee

\be \beta_{\tilde{\zeta}}^{(1)}= \tilde{\zeta}(
\bar{\gamma}_{\Phi}^{(1)}+ \bar{\gamma}_{\Sigma}^{(1)}+
\bar{\gamma}_{\Theta}^{(1)})+2\zeta ( \tilde{\gamma}_{\Phi}^{(1)}+
\tilde{\gamma}_{\Sigma}^{(1)}+ \tilde{\gamma}_{\Theta}^{(1)}) -70
g_{10}^2(\tilde{\zeta}-2 M \zeta) \ee

\be \beta_{\tilde{\bar{\zeta}}}^{(1)}= \tilde{\bar{\zeta}}(
\bar{\gamma}_{\Phi}^{(1)}+ \bar{\gamma}_{\bar{\Sigma}}^{(1)}+
\bar{\gamma}_{\Theta}^{(1)})+2\bar{\zeta} (
\tilde{\gamma}_{\Phi}^{(1)}+ \tilde{\gamma}_{\bar{\Sigma}}^{(1)}+
\tilde{\gamma}_{\Theta}^{(1)}) -70 g_{10}^2(\tilde{\bar{\zeta}}-2
M \bar{\zeta}) \ee

\be
\beta_{\tilde{h}}^{(1)}=\bar\gamma_{H}^{(1)}\tilde{h}+\tilde{h}.\bar\gamma_{\Psi}^{(1)}
+(\bar\gamma_{\Psi}^{(1)})^T.\tilde{h}+2\tilde{\gamma}_{H}^{(1)} h
+2 (h.\tilde{\gamma}_{\Psi}^{(1)} +
(\tilde{\gamma}_{\Psi}^{(1)})^T.h)-\frac{63}{2} g_{10}^2
(\tilde{h}-2 M h) \ee

\be \beta_{\tilde{g}}^{(1)}=\bar\gamma_{\Theta}^{(1)}\tilde{g}-
\tilde{g}.\bar\gamma_{\Psi}^{(1)} +(\bar\gamma_{\Psi}^{(1)})^T
.\tilde{g}+2\tilde{\gamma}_{\Theta}^{(1)}.g + 2(g.
\tilde{\gamma}_{\Psi}^{(1)} -(\tilde{\gamma}_{\Psi}^{(1)} )^T
.g)-\frac{87}{2} g_{10}^2 (\tilde{g}-2 M g) \ee

\be
\beta_{\tilde{f}}^{(1)}=\bar\gamma_{\bar{\Sigma}}^{(1)}\tilde{f}+\tilde{f}.\bar\gamma_{\Psi}^{(1)}
+ (\bar\gamma_{\Psi}^{(1)} )^T
.\tilde{f}+2\tilde{\gamma}_{\bar{\Sigma}}^{(1)}.f +2 (f
.\tilde{\gamma}_{\Psi}^{(1)} + (\tilde{\gamma}_{\Psi}^{(1)})^T
.f)-\frac{95}{2} g_{10}^2 (\tilde{f}-2 M f) \ee

\be \beta_{b_{\Phi}}^{(1)}=2 b_{\Phi} \bar{\gamma}_{\Phi}^{(1)}+4
\mu_{\Phi} \tilde{\gamma}_{\Phi}^{(1)}-48 g_{10}^2(b_{\Phi}-2 M
\mu_{\Phi})\ee

\be \beta_{b_{H}}^{(1)}=2 b_{H} \bar{\gamma}_{H}^{(1)}+4 \mu_{H}
\tilde{\gamma}_{H}^{(1)}-18 g_{10}^2(b_H-2 M \mu_H)\ee

\be \beta_{b_{\Theta}}^{(1)}=2 b_{\Theta}
\bar{\gamma}_{\Theta}^{(1)}+4 \mu_{\Theta}
\tilde{\gamma}_{\Theta}^{(1)}-42 g_{10}^2(b_{\Theta}-2 M
\mu_{\Theta})\ee

\be \beta_{b_{\Sigma}}^{(1)}= b_{\Sigma}(
\bar{\gamma}_{\Sigma}^{(1)}+\bar{\gamma}_{\bar{\Sigma}}^{(1)})+ 2
\mu_{\Sigma}(
\tilde{\gamma}_{\Sigma}^{(1)}+\tilde{\gamma}_{\bar{\Sigma}}^{(1)})-50
g_{10}^2(b_{\Sigma}-2 M \mu_{\Sigma})\ee

\bea \beta_{m^2_{\Phi}}^{(1)}&=&2 \bar{\gamma}_{\Phi}^{(1)}
m^2_{\Phi}+720 m^2_{\Phi}|\lambda|^2+
m^2_{H}(12|\gamma|^2+12|\bar{\gamma}|^2+8|\kappa|^2)\nonumber\\&&
+m^2_{\Theta}(8|\rho|^2+120(|\zeta|^2+|\bar{\zeta}|^2)+8|\kappa|^2)
+m^2_{\Sigma}(480|\eta|^2+12|\gamma|^2+120|\zeta|^2)\nonumber\\&&+m^2_{\bar{\Sigma}}(480|\eta|^2
+12|\bar{\gamma}|^2+120|\bar{\zeta}|^2)
+2\hat{\gamma}_{\Phi}^{(1)}-96|M|^2 g_{10}^2\eea

\bea \beta_{m^2_{H}}^{(1)}&=&2
\bar{\gamma}_{H}^{(1)}m^2_{H}+m^2_{\Phi}(252(|\gamma|^2+|\bar{\gamma}|^2)+168|\kappa|^2)
+168m^2_{\Theta}|\kappa|^2+252m^2_{\Sigma}|\gamma|^2\nonumber\\&&+252m^2_{\bar{\Sigma}}|\bar{\gamma}|^2
+2\hat{\gamma}_{H}^{(1)}-36|M|^2 g_{10}^2+32 \text
Tr[h^{\dag}.m^2_{\tilde{\Psi}}.h]\eea

\bea \beta_{m^2_{\Theta}}^{(1)}&=&2
\bar{\gamma}_{\Theta}^{(1)}m^2_{\Theta}+m^2_{\Phi}(14(|\kappa|^2+|\rho|^2)+210(|\zeta|^2+|\bar{\zeta}|^2))
+14m^2_{\Theta}|\rho|^2+14m^2_{H}|\kappa|^2\nonumber\\&&+210m^2_{\Sigma}|\zeta|^2
+210m^2_{\bar{\Sigma}}|\bar{\zeta}|^2
+2\hat{\gamma}_{\Theta}^{(1)}-84|M|^2 g_{10}^2+32 \text
Tr[g^{\dag}.m^2_{\tilde{\Psi}}.g]\eea

\bea \beta_{m^2_{\Sigma}}^{(1)}&=&2
\bar{\gamma}_{\Sigma}^{(1)}m^2_{\Sigma}+m^2_{\Phi}(400|\eta|^2+20|\gamma|^2+200|\zeta|^2)
+200m^2_{\Theta}|\zeta|^2+20m^2_{H}|\gamma|^2\nonumber\\&&+400m^2_{\bar{\Sigma}}|\eta|^2
+2\hat{\gamma}_{\Sigma}^{(1)}-100|M|^2 g_{10}^2\eea

\bea \beta_{m^2_{\bar{\Sigma}}}^{(1)}&=&2
\bar{\gamma}_{\bar{\Sigma}}^{(1)}m^2_{\bar{\Sigma}}+m^2_{\Phi}(400|\eta|^2+20|\bar{\gamma}|^2+200|\bar{\zeta}|^2)
+200m^2_{\Theta}|\bar{\zeta}|^2+20m^2_{H}|\bar{\gamma}|^2\nonumber\\&&+400m^2_{\Sigma}|\eta|^2
+2\hat{\gamma}_{\bar{\Sigma}}^{(1)}-100|M|^2 g^2_{10}+128 \text
Tr[f^{\dag}.m^2_{\tilde{\Psi}}.f]\eea

\bea \beta_{m^2_{\tilde{\Psi}}}^{(1)}&=&\bar{\gamma}_\Psi^{(1)}
.m^2_{\tilde\Psi}+ m^2_{\tilde\Psi}.\bar{\gamma}_\Psi^{(1)} +10
h^{\dag}.m^2_{\tilde{\Psi}}.h+120
g^{\dag}.m^2_{\tilde{\Psi}}.g+252 f^{\dag}.m^2_{\tilde{\Psi}}.f+10
m^2_{H} h^{\dag}.h\nonumber\\&&+120
m^2_{\Theta}g^{\dag}.g+252m^2_{\bar{\Sigma}}
f^{\dag}.f+2\hat{\gamma}_{\Psi}^{(1)}-45|M|^2 g^2_{10}\eea

\end{document}